\documentclass[usegraphicx,usenatbib,useAMS]{mn2e}

\usepackage{times} 
\usepackage{aas_macros}     
\usepackage{amssymb}        
\usepackage{xspace}

\newcommand{\bc}{\begin{center}}
\newcommand{\ec}{\end{center}}

\newcommand{\gsim}{\gtrsim} 
\newcommand{\lsim}{\lesssim} 

\newcommand{\Msol}{M_\odot}
\newcommand{\hMsol}{h^{-1}{\rm M_\odot}}
\newcommand{\Lsol}{L_\odot}
\newcommand{\hLsol}{h^{-2}{\rm L_\odot}}

\newcommand{\ergs}{{\rm erg\,s^{-1}}}

\newcommand{\mum}{\mu{\rm m}}

\newcommand{\Mpc}{{\rm Mpc}}

\newcommand{\kms}{{\rm km\,s^{-1}}}
\newcommand{\yr}{{\rm yr}}

\newcommand{\mJy}{{\rm mJy}}

\newcommand{\WHz}{{\rm W Hz^{-1}}}

\newcommand{\GALFORM}{\textsc{galform}\xspace}

\newcommand{\Herschel}{{\em Herschel}\xspace}
\newcommand{\Spitzer}{{\em Spitzer}}
\newcommand{\IRAS}{{\em IRAS}}

\newcommand{\ISO}{{\em ISO}}
\newcommand{\SCUBA}{{\em SCUBA}}

\newcommand{\AKARI}{{\em AKARI}}
\newcommand{\GALEX}{{\em GALEX}}

\newcommand{\WISE}{{\em WISE}}

\newcommand{\ASKAP}{{\em ASKAP}}
\newcommand{\GMRT}{{\em GMRT}}

\newcommand{\HATLAS}{\textsc{h-atlas}\xspace}
\newcommand{\GAMA}{\textsc{gama}\xspace}
\newcommand{\SDSS}{\textsc{sdss}\xspace}
\newcommand{\UKIDSS}{\textsc{ukidss}\xspace}

\title[Far-IR luminosity function of galaxy groups]
{{\it Herschel-ATLAS/GAMA}:How does the far-IR luminosity
  function depend on galaxy group properties?}

\author[Guo et al.]
       { \parbox{18cm}{Qi Guo$^{1,2}$\thanks{Email:qi.guo2010@gmail.com}, Cedric Lacey$^1$, 
	Peder Norberg$^1$, Shaun Cole$^1$,  Carlton Baugh$^1$, Carlos Frenk$^{1}$, 
        Asantha Cooray$^3$, Simon Dye$^4$, 
	N. Bourne$^4$, L. Dunne$^5$,  S. Eales$^6$, R.J. Ivison$^7$, S.J. Maddox$^5$, 
	M. Alpasan$^8$, I. Baldry$^9$, J. Bland-Hawthorn,$^{10}$, S.P. Driver$^{8,11}$, 
	A. Robotham$^{11}$
	 }	
    \\     
    \\
    $^1$ Institute for Computational Cosmology, Department of Physics, University of Durham, South Road, Durham, DH1 3LE, UK \\ 
    $^2$ Partner Group of the Max-Planck-Institut f\"ur
    Astrophysik, National Astronomical Observatories, Chinese
    Academy of Sciences, \\Beijing, 100012, China \\
    $^3$ Dept. of Physics \& Astronomy, University of California, Irvine, CA,92697, USA     \\
    $^4$ School of Physics and Astronomy, University of Nottingham, Nottingham NG7 2RD, UK \\
    $^5$ Department of Physics and Astronomy, University of Canterbury, Private Bag 4800, Christchurch, New Zealand\\
    $^6$ School of Physics \& Astronomy, Cardiff University, The Parade, Cardiff, CF24 3AA, UK\\  
    $^7$ UK Astronomy Technology Centre, Royal Observatory, Blackford Hill, Edinburgh EH9 3HJ, UK\\
    $^8$ Scottish Universities Physics Alliance (SUPA), School of Physics and Astronomy, University of St Andrews, North Haugh, St Andrews KY16 9SS, UK \\
    $^9$ Astrophysics Research Institute, Liverpool John Moores
       University, IC2, Liverpool Science Park, 146 Brownlow Hill,
       Liverpool, L3 5RF \\
    $^{10}$ Sydney Institute for Astronomy, University of Sydney, NSW
       2006, Australia \\
    $^{11}$ International Centre for Radio Astronomy Research (ICRAR), University of Western Australia, Crawley, WA 6009, Australia 
       }
\begin{document}

\maketitle

\begin{abstract}
  We use the Herschel  ATLAS (\HATLAS) Phase I data to study the conditional
  luminosity function of far-IR (250 $\mu$m) selected galaxies in
  optically-selected galaxy groups from the \GAMA spectroscopic survey, as
  well as environmental effects on the far-IR-to-optical colour. We
  applied two methods, which gave consistent results for the far-IR
  conditional luminosity functions. The direct matching method matches
  \HATLAS sources to \GAMA/\SDSS galaxies, then links the optical
  counterparts to \GAMA groups. The stacking method counts the number
  of far-IR sources within the projected radii of \GAMA groups,
  subtracting the local background. We investigated the dependence of
  the far-IR ($250 \mum$) luminosity function on group mass in the
  range $10^{12} < M_{\rm h} < 10^{14} \hMsol$ and on redshift in the range
  $0<z<0.4$, using a sample of 3000 groups containing \HATLAS sources
  with \GAMA redshifts over an area of 126 deg$^2$. We find that the characteristic $250 \mum$
  luminosity, $L^{\ast}(250)$, increases with group mass up to $M_{\rm h} \sim
  10^{13} \hMsol$, but is roughly constant above this, while it
  increases with redshift at high group masses, but less so at low
  masses. We also find that the group far-IR luminosity-to-mass ratio
  $L(250)/M_{\rm h}$ increases with redshift and is higher in low-mass
  groups.
  We estimate that around 70\% of the $250 \mum$ luminosity density in
  the local universe is contributed by groups with $M_{\rm h} > 10^{12}
  \hMsol$. We also find that the far-IR-to-optical colours of \HATLAS
  galaxies are independent of group mass over the range $10^{12} < M_{\rm h}
  < 10^{14} \hMsol$ in the local universe.  We also compare our
  observational results with recent semi-analytical models, and find
  that none of these galaxy formation model can reproduce the conditional
  far-IR luminosity functions of galaxy groups.
\end{abstract}


\section{Introduction}
Star formation is one of the most important processes determining the
formation and evolution of the galaxies. Theoretical work
 suggests that {\em in situ} star
formation  dominates over the accretion and mergers 
of satellite galaxies for the growth in stellar mass of
galaxies less massive than the Milky Way at all redshifts \citep[e.g.][] {Guo2008, Parry2009}. Even at the
Milky Way mass, star formation is the primary means of adding stellar mass at $z \gsim 1$. Observational
studies have measured star formation rates (SFRs) from the local
Universe to high redshifts. A picture in which the overall star formation
density increases with redshift and peaks at around $z \sim 2$ has now
been well established \citep[e.g.][]{Madau1998, Hopkins2007}.  Methods
to infer the SFR include the direct measurement of the rest-frame UV
luminosity \citep[e.g.][]{Lilly1996, Madau1998,
Steidel1999,Salim2007}, or emission lines such as H$\alpha$ and [OII] emission lines
\citep[e.g.][]{Gallego1995, Brinchmann2004,Sobral2011}, all of which
trace massive young stars. However, these methods are
subject to uncertain corrections for dust extinction, which varies in regions of different local
properties, as well as depending on the inclination of the galaxy. UV
photons heat the dust around star forming regions and are then
reprocessed by the dust and their energy is re-emitted in the mid- and
far-IR range, with the dust emission typically peaking at a wavelength
around 100$\mu$m. About half of the starlight is absorbed and
re-emitted  over the history of the Universe \citep{Puget1996,Hauser1998}
(some studies show that an even larger fraction of the UV light is
reprocessed, e.g. \citep{Buat2007}). Observations at IR wavelengths
are thus an essential complement to UV and optical tracers of star
formation. Previous surveys in the IR include that by \IRAS, which
measured the far-IR emission at $\leq 100 \mu$m, which mainly
constrains the emission from warm dust in bright galaxies
\citep{Dunne2001}, while more recent surveys of dust emission focused
either on the mid-IR (\ISO, \Spitzer) or sub-mm (e.g. \SCUBA)
wavelengths, and therefor misses the peak in the dust emission, and hence
requires uncertain extrapolations to infer total IR luminosities.
\Herschel\footnote{Herschel is an ESA space observatory with science
    instruments provided by European-led Principal Investigator
    consortia and with important participation from NASA.}
    \citep{Pilbratt2010} observations span the far-IR wavelengths
    $60-700~\mu$m, covering the peak of the dust emission from
    star-forming galaxies. Moreover, as the largest open-time key
    project on \Herschel, the Herschel Astrophysical Terahertz Large
    Area Survey (\HATLAS) \citep{Eales2010} provides far-IR imaging
    and photometry over an area of 550 deg$^2$, in five channels centred
    on 100, 160, 250, 350, and 500 $\mu$m, ideal for using the far-IR
    emission to estimate the dust obscured star formation rate.

The star formation rate of galaxies depends on 
stellar mass, redshift, and environment. It has been known for
many years that the fraction of star forming galaxies decreases as the
mass of the host dark matter halo increases, from isolated field
galaxies up to rich clusters
\citep[e.g.][]{Dressler1980,Kimm2009}. The fraction of actively
star-forming galaxies in groups and clusters also increases with
redshift \citep[e.g.][]{Butcher1978}. However, focusing only on the
population of star-forming galaxies, the effect of galaxy environment
on star formation activity is still under debate. Most studies find no
dependence of the SFR of star-forming galaxies at a given stellar
mass on group/cluster environment or local density, from low ($z=0$)
to intermediate ($z<0.5$) redshifts \citep{Balogh2004, Tanaka2004,
Weinmann2006, Peng2010, McGee2011}. This independence has also been
found at high redshift ($z \sim 1$) \citep{Ideue2012}. However, some
other studies conflict with this conclusion \citep{Lewis2002,
Gomez2003, Welikala2008}, suggesting that galaxy SFRs are more
strongly suppressed in highly overdense regions.

Most previous work on the dependence of galactic SFRs on environment
has used the UV continuum or the H$\alpha$ emission to estimate
SFRs. In this paper, we revisit this problem by looking at an
important tracer of the dust-obscured SFR, the far-IR emission. Early
work on the IR properties of galaxies in rich clusters based on {\IRAS}
 and {\ISO}  observations is reviewed by \citet{Metcalfe2005}. There have
been several studies using mid-IR observations, mainly the Spitzer
24$\mum$ band, to estimate the IR luminosity functions (LFs) of galaxy
clusters ($M \gsim 10^{14} \Msol$)
\citep[e.g][]{Bai2006,Bai2009,Chung2010,Finn2010,Goto2010,Biviano2011},
and one measurement of the IR LF of massive galaxy groups ($10^{13}
\lsim M \lsim 10^{14} \Msol$) \citep{Tran2009}. However these studies had the
drawback that they had to extrapolate in wavelength in order to
estimate total IR luminosities. SFRs estimated from mid-IR
luminosities have been used to study the fraction of star-forming
galaxies in different density environments
\citep[e.g.][]{Koyama2008,Tran2009}, and also to study the dependence
of the specific star formation rate (sSFR), defined as the ratio,
$SFR/M_{\star}$, of SFR to stellar mass $M_{\star}$, on local density
and group or cluster environment
\cite[e.g.][]{Elbaz2007,Bai2010}. Mid-IR observations have also been
used to estimate the evolution of the ratio $L_{\rm IR}/M_{\rm h}$ of total IR
luminosity to dark matter halo mass $M_{\rm h}$ for rich clusters
\citep{Geach2006, Koyama2010, Webb2013}. These studies have recently
been extended to the far-IR and to galaxy groups by
\citet{Popesso2012}, who used \Herschel observations to measure
$L_{\rm IR}/M_{\rm h}$ for 9 rich clusters ($M \sim 10^{15} \Msol$) and 9 groups
($M \sim 5\times 10^{13} \Msol$) at redshifts $0.1 \lsim z \lsim 1$.


In this paper, we directly measure the far-IR LFs and $L_{\rm IR}/M_{\rm h}$
ratios of a very large sample ($\sim 3000$) of galaxy groups and
clusters covering a wide range in mass, $10^{12} < M_{\rm h} < 10^{14}
\hMsol$, in the low-redshift $z<0.4$ Universe using data from
\Herschel. We also use our sample to measure the dependence of the
dust-obscured sSFR on group mass. The galaxy groups are optically
selected from the \GAMA spectroscopic survey \citep{Driver2009}. Our study has several
advantage over previous studies of the same range of group mass and
redshift: (a) We use far-IR observations, which provide a much more
robust measure of the total IR luminosity, and hence of the
dust-obscured SFR, than is possible using mid-IR data. (b) We probe a much larger range
of group mass than was available to previous studies, which were restricted to quite
massive groups, $10^{13} \lsim M_{\rm h} \lsim 10^{14} \hMsol$. (c) We have
a much larger sample of groups than previous studies, which had
samples of $\sim 10$ groups at most. (d) Our group sample, being
optically selected, is much more complete than the X-ray selected
samples used in many previous IR studies. (e) We study the IR LF of
groups down to $L_{\rm IR} \sim 10^{9} \Lsol$, much fainter than most previous
studies of groups and clusters, which were restricted to $L_{\rm IR} \sim
10^{10} \Lsol$ or brighter.



The first step in our study is to measure the galaxy abundance in
groups and clusters as a function of their far-IR
luminosities. Similar techniques have been developed extensively in
the optical range \citep[e.g.][]{Jing1998, Berlind2002,Yang2003}, while the
far-IR is almost unexplored due to the previous lack of deep and sufficiently large
surveys at these wavelengths. The \HATLAS is a perfect survey for this
study. To identify galaxy groups, we use group catalogues
\citep{Robotham2011} based on an optical redshift survey -- the Galaxy
And Mass Assembly I survey \citep[\GAMA I,][]{Baldry2010, Driver2009, Driver2011,
Hill2011, Taylor2011, Kelvin2012}. The abundance of far-IR galaxies
within a given \GAMA group is then measured using two methods. One
is to match \HATLAS sources to \GAMA galaxies \citep{Smith2011},
calculating the abundance of the far-IR-detected group members
directly. The other is to calculate the abundance of \HATLAS sources
within a projected radius around the group centre after subtracting
the contribution from the background. After measuring the far-IR
conditional luminosity function (CLF) for groups of different masses and
redshifts, we further study its properties, including
the characteristic far-IR luminosity $L^{\ast}$ and luminosity-to-mass
ratio, and their correlation with the masses and redshifts of the host
groups. The first method also enables us to study the variation of the
far-IR-to-optical colour (which is an indicator of the specific star
formation rate) in the field, groups and clusters.

This paper is organized as follows. In \S\ref{sec:data} we briefly
describe the two catalogues used in this work: \HATLAS Phase I and
\GAMA I (including groups). The two methods used to count group
members, as well as the data description are also presented in
\S\ref{sec:data}. The far-IR luminosity functions in groups of
different masses and redshifts are presented in \S\ref{sec:LFs}. In
this section, we also discuss the relationship between the total far-IR
luminosity and group mass and its evolution with redshift. In
\S\ref{sec:colours}, we discuss the far-IR - optical colour, focusing
on environmental effects and redshift evolution. A comparison with
predictions from galaxy formation models is presented in
\S\ref{sec:models}.  Our main results are summarized in  \S\ref{sec:conc}.

Throughout this paper we assume a flat $\Lambda$CDM cosmology with
$\Omega_{\rm m}$ = 0.25, $\Omega_{\Lambda}$ = 0.75, h = 0.73  where $H_0 = 100 
\kms\Mpc^{-1}$ and power spectrum normalization $\sigma_8 =0.9$. 
This power spectrum normalization is only relevant for our model predictions as presented in Sec.\ref{sec:models}.

\section{Data and Methods}
\label{sec:data}
In \S\ref{sec:gamadata} we describe the GAMA data, in \S\ref{sec:hatlasdata} the Herschel-Atlas data, and in \S\ref{sec:CLF_methods} 
we outline how we measure the luminosity of groups. 
\subsection{\GAMA I}
\label{sec:gamadata}
The \GAMA I survey is an optical spectroscopic galaxy survey covering
142 deg$^2$ in three equal-sized regions on the celestial equator, to
apparent $r$-band magnitude $r_{AB} = 19.4$ in two regions (G09 and
G15) and $r_{AB} = 19.8$ in one region (G12).\footnote{The \GAMA II survey
 reaches $r_{AB} = 19.8$ in all
fields. However, the group catalogue we use is based on \GAMA I. In
order to have a self-consistent analysis, we adopt the \GAMA I data
release rather than the \GAMA II survey} It
contains 110,192 galaxies with optical/near-IR imaging
(from \SDSS, \UKIDSS, \textsc{kids}, \textsc{viking}, with the latter two still
underway), and complementary observations from the UV (\GALEX) through
to the mid and far-IR (\WISE, \Herschel) and the radio (\ASKAP, \GMRT,
underway).  The redshift completeness to $r$-band magnitude 19.4 reaches
98.7\% \citep{Driver2011}.  To
simplify the selection function, we limit ourselves to $r<19.4$. This
leads to a sample of 93325 galaxies, with a redshift
coverage of $0<z<0.5$ centred at around $z \sim 0.2$.

 Using the GAMA I optically selected redshift catalogue, \cite{Robotham2011} 
used a redshift space friends-of-friends (FoF) grouping algorithm
to create the GAMA I group catalogue. Systems with 2
or more optical members are identified as galaxy groups. In total,
there are 12.2k GAMA groups, and around 34\% of GAMA galaxies
belong to groups. Total group masses used in this study
are inferred from the total r-band luminosity of the group, its redshift
and the group multiplicity, following the description given in
\cite{Robotham2011} and implemented in Han et al. (in prep).
The GAMA group catalogue has been extensively tested against a
set of mock GAMA lightcones, following the method described in
\cite{Merson2013}. In summary, the mocks are constructed from the
Millennium  $\Lambda$CDM dark matter N-body simulation
\citep{Springel2005}, populated with galaxies using the GALFORM
semi-analytical galaxy formation model \citep{Cole2000}, using the
\cite{Bower2006} model as input. Finally, the raw GALFORM
lightcones are abundance matched to precisely reproduce the
GAMA r-band luminosity function \citep{Loveday2012}, resulting
in minor modifications to the r-band magnitudes (typically less
than 0.1 mag). This is consistent with differences expected to arise
from different magnitude definitions, which are not included
in the lightcone pipeline. Readers are referred to
\cite{Merson2013} and \cite{Robotham2011} for further details
on the mocks, and in particular to the latter for a list of known
limitations specific to the GAMA lightcone mocks.


For the present study, the completeness of the group catalogue
as function of group mass and redshift needs to be addressed. Using the mocks, we estimate this completeness to
be about 90\% for $z<$0.2 and M$_h>$10$^{13}$ M$_{\odot}$/h,while it decreases
strongly with decreasing group mass and  increasing redshift to
below 20\%  for e.g. groups less massive than 10$^{13}$ M$_{\odot}$/h in the
redshift range 0.1$<z<$0.3 .
These completenesses correspond to upper limits, as they do
not account for uncertainties in the group mass estimate,
nor in the grouping. A comprehensive investigation, including
uncertainties from applying the group finder to a different set of mocks,
is currently underway and beyond the scope of the present paper.
We note here that the underlying assumption for the rest of the paper
is that the identified groups of a given mass are an unbiased sample of all
groups of that mass.


\subsection{\HATLAS Phase I}
\label{sec:hatlasdata}
The \HATLAS Phase~I Data Release consists of three equatorial fields
(G09, G12 and G15), covering 135~deg$^2$ in total. The overlap between
\HATLAS and the \GAMA-I survey is about 126~deg$^2$. \HATLAS has
imaging in five far-IR bands centred on 100, 160, 250, 350 and 500
$\mu$m, using the PACS \citep{Poglitsch2010} and SPIRE
\citep{Griffin2010} instruments. The median values of the 5$\sigma$ flux limits
are 132, 126, 32, 36, and 45~mJy respectively for the five
wavelengths.  There are 78.0k sources brighter than the 5$\sigma$
detection limit in one or more of the 3 SPIRE bands \citep{Rigby2011}.
In this paper, we work with a 250~$\mu$m flux-limited sample, since
this is the most sensitive band and has the best positional accuracy
of the 3 SPIRE bands. 

  In the \HATLAS Phase I Data Release, sources brighter than the
  5$\sigma$ flux limit at 250~$\mu$m have been matched to the $r$-band
  selected ($r < 22.4$) \SDSS galaxy imaging catalogue using a
  likelihood ratio method \citep{Sutherland1992}. The application of
  the method to the \HATLAS survey is described in detail in
  \citet{Smith2011}, but we give a brief description here.  For a
  potential optical counterpart with $r$-band apparent magnitude $m$
  at angular distance $r$ from the estimated position of the \HATLAS
  source, the likelihood ratio $L$ is calculated as $L =
  \frac{q(m)f(r)}{n(m)}$, and gives the ratio of the probability that
  the optical source is the correct ID to the corresponding
  probability that it is an unrelated background source.  The
  positional errors for the \HATLAS sources, which determine the
  radial probability distribution {\it f(r)}, are determined using
  histograms of the separations between the positions of 250~$\mu$m
  sources and those of galaxies in the SDSS DR7 r-band catalogue
  within 50 arcsec.
  {\it n(m)} is the probability that a background SDSS source is
  observed with magnitude {\it m}, which is well-defined. {\it q(m)}
  is the probability for a true counterpart to a 250 $\mu$m source to
  have a magnitude {\it m}, which is calculated as the normalized
  magnitude distribution of the SDSS sources within 10 arcsec of each
  250 $\mu$m source after subtracting the background, multiplied by
  the fraction $Q_0$ of true counterparts which are above the SDSS
  limit. \citet{Smith2011} measure $Q_0=0.59$. A reliability value
  ($R_{\rm LR} $) is then assigned to each potential optical
  counterpart (hereafter candidate), which allows for the presence of
  other optical candidates. $R_{\rm LR} $ is the Bayesian probability
  that the candidate is the true counterpart.
  Following \citeauthor{Smith2011}, we take the threshold for a
  reliable match as $R_{\rm LR} >0.8$, for which choice they estimate
  that 96\% of the assigned IDs are the true optical counterparts.

For this paper, we use the sample of 66.2k \HATLAS sources with
250~$\mu$m flux above $35~\mJy$, which is higher than the median
5$\sigma$ value to guarantee a uniform selection. There are then
29.8k candidate optical counterparts with $R_{\rm LR} >0.8$ (45\% of
our total 250~$\mu$m \HATLAS sample).  Out of this matched sample,
24.2k galaxies are in the area overlapping with the \GAMA-I
survey. Applying a uniform cut of $r<19.4$ in all three regions
overlapping with \GAMA-I leads to 10.5k \HATLAS galaxies with reliable
counterparts in the \GAMA spectroscopic sample, corresponding to 43\%
of the sources with an optical counterpart in \SDSS in the \GAMA-I
overlap region. This sample with spectroscopic redshifts forms the
direct matching catalogue which we use in most of this work.

  In this paper, we study \HATLAS galaxies at $z<0.4$. An important
  question is whether a significant fraction of such galaxies have
  optical counterparts which are too faint to appear in the SDSS
  imaging catalogue. To answer this question, we use data from the
  Herschel HerMES survey \citep{Oliver2012} of the COSMOS field, which
  also has very deep optical, 3.6~$\mum$ and 24~$\mum$ imaging
  data. The HerMES team find optical counterparts to Herschel sources
  by first matching Herschel sources to 24~$\mum$ sources
  \citep{Roseboom2010,Roseboom2012}, which method is estimated to be
  highly complete for this field at the fluxes of interest here
  ($S(250\mum) > 35$mJy). The 24~$\mum$ sources are then matched to
  the closest 3.6~$\mum$ source within 2~arcsec, and each 3.6~$\mum$
  source is matched to the nearest optical source within 1~arcsec.
  They then obtain redshifts for these sources from the COSMOS
  photometric redshift catalogue \citep{Ilbert2009}. In this way, they
  measure a highly complete redshift distribution for Herschel sources
  with $S(250\mum) > 35$mJy. From this catalogue, they find that all
  sources with $S(250\mum) > 35$mJy and having optical counterparts
  fainter than $r>22.4$ lie at redshifts $z>0.4$ (Lingyu Wang, private
  communication).

There are 10.7k \GAMA groups in the \HATLAS overlap region, of which
3.0k groups have mass $10^{12} < M_{\rm h} < 10^{14} \hMsol$ and
redshifts in the range $0<z<0.4$ and contain 1 or more reliable
$r<19.4$ counterparts in \HATLAS. In total there are 3.2k
\HATLAS-\GAMA galaxies in \GAMA groups.

\begin{figure}
\bc
\hspace{-0.6cm}
\includegraphics[width=8.5cm]{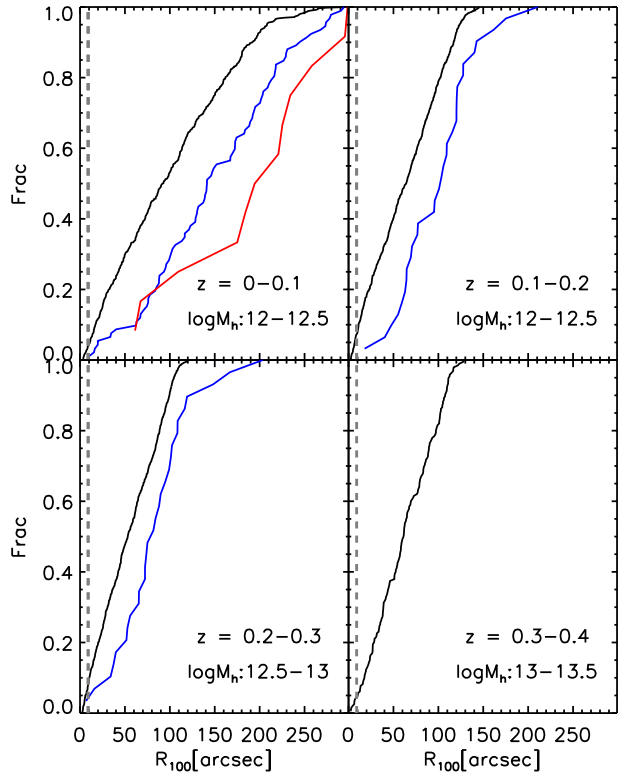}
\caption{Cumulative distribution of groups as a function of
    $R_{100}$, the angular radius of the most distant group member
    from the group centre. Redshift ranges and mass ranges for the
    groups are indicated in the bottom right corner of each
    panel. Black, blue and red curves are for groups with optical
    multiplicity 2, 3 and $>3$, respectively.The vertical grey dashed
    curves in each panel indicate the Herschel beam radius of 9~arcsec
    at 250~$\mum$.}
\label{fig:size}
\ec
\end{figure}

  Note that the FWHM beamsize of Herschel is about 18~arcsec at
  250~$\mum$, which could potentally lead to source confusion,
  especially in high density regions like groups. The general effects
  of confusion in the \HATLAS survey were found by \citet{Rigby2011}
  to be modest at 250~$\mum$. In Fig.~\ref{fig:size} we show the size
  distribution of the GAMA groups in different ranges of redshift and
  mass, compared to the beamsize of Herschel at 250~$\mu$m. Since the
  group size is an increasing function of group mass, at each redshift
  we only present the lowest group mass range which contains \HATLAS
  galaxies. The figure shows that only a few percent of the groups in our
  sample have angular sizes comparable to or smaller than the beam
  radius of Herschel at 250~$\mum$. We have also visually checked the
  spatial distribution of \GAMA galaxies in a random subset of groups,
  and find that in most cases, the separations between optical members
  are much larger than the beamsize at 250$\mu$m. We therefore
  conclude that the effects of source confusion due to other group
  members should be very small for this study.

To k-correct the observed 250~$\mu$m flux to the rest~frame 250$\mu$m
luminosity, we assume the dust emission SED is a modified black-body,
as in \cite{Guo2011}:
\begin{equation}
L_{\nu} \propto B_{\nu}(T)\nu^{\beta},
\label{eq:kcorrection}
\end{equation}
where $B_{\nu}(T)$ is the Planck function, and we assume $\beta$ = 1.5 and fit the
temperature using as many far-IR bands as are detected.  We then calculate the k-correction
for each source using its individually estimated temperature.

The median temperature of our sample
is 26~K. For this
assumed SED shape and temperature, the ratio of the total IR luminosity $L_{\rm IR}$ (integrated
over 8--1000~$\mu$m) to $\nu\rm L_{\nu}$(250, rest frame) is 5.5 for the median temperature, so we have the
conversion between monochromatic and total IR luminosities of
\begin{equation}
L_{\rm IR}/{\rm \Lsol} = 1.8\times 10^{10}  \, 
L_{\nu}(250\mum)/(10^{24} \WHz) .
\label{eq:totL}
\end{equation}
 
To relate far-IR luminosities to dust-obscured SFRs, we use the
relation derived by \citet{Kennicutt1998b}, multiplied by $0.63$ which
corresponds to a \citet{Chabrier2003} IMF over a stellar mass range
$0.1 < m <100 {\rm \Msol}$. This gives
\begin{eqnarray}
SFR/({\rm\Msol}\yr^{-1}) &=& 2.8 \times 10^{-44}
L_{\rm IR}/(\ergs) \nonumber \\
&=& 1.8 \, L_{\nu}(250\mum)/(10^{24} \WHz) ,
\label{eq:SFR-LIR}
\end{eqnarray}
where in the second line we have assumed the SED shape described
above with T = 26~K.

The conversation from L$_{250\mum}$ to the SFR above is only to
illustrate the typical SFR. In practice, when calculating the SFR in
Sec. ~\ref{sec:colours}, we use the temperature for each source from
fitting its own SED. Note that L$_{\rm IR}$ could be underestimated
with the assumption of a single modified black body for the SED
fitting, given that the SED of dust emission could be more complicated
and there are contributions from hot dust, PAH and VSG which emit in
the mid-IR. The true total emission from 8-1000 $\mu$m could be higher
by around 30-50\% than that obtained using Eq. (\ref{eq:totL}). On the
other hand, since a lot of the dust in H-ATLAS sources seem to be
heated by older stars and not recent SFR, the conversation from the
total L$_{\rm IR}$ in Eq. (\ref{eq:SFR-LIR}) could overestimate the
derived SFR. These two effects partly compensate each other and our
results should be robust within a factor of 2.

\subsection{Methods for measuring the far-IR LF of galaxy groups}
\label{sec:CLF_methods}
We use two independent methods to estimate the abundance of
group/cluster galaxies as a function of their far-IR luminosity.
\subsubsection{Direct matching method}
 Our
first method (hereafter, the direct method) uses the matched
\HATLAS-\GAMA galaxy catalogue. The \GAMA group catalogue
\citep{Robotham2011} lists the \GAMA galaxies in each \GAMA
group/cluster. The direct matching catalogue, on the other hand,
establishes the link between the \HATLAS source and the $r$-band
selected \GAMA galaxy. The combination of these two directly links the
\GAMA group/cluster with its \HATLAS members. Around 33\% of the
galaxies in the matched \HATLAS-\GAMA catalogue are identified as
group members, the same as the corresponding fraction of \GAMA
galaxies (34\%).

The mean far-IR luminosity function of galaxies in groups in a certain
mass and redshift range is calculated using
\begin{equation}
\Phi(L_i) \, \Delta\log L_i  
= \frac{\Sigma_{j = 0}^{N_{\rm group}}n_{i,j}}{\Sigma_{j = 0}^{N_{\rm group}} N_{i,j}} ,
\end{equation}
where $\Phi(L_i) \, \Delta\log L_i$ is the number of galaxies per
group in the $i$th luminosity bin $L_i$, $n_{i,j}$ is the number of
 matched galaxies in the $i^{th}$ luminosity bin for the $j$th group,
and $N_{\rm group}$ is the total number of groups for a given redshift bin
and group mass bin. The factor $N_{i,j}$ specifies whether the $j^{th}$
group contributes to the measurement in the $i$th luminosity bin,
given the far-IR flux limit and redshift, and is defined as $N_{i,j} $
= 1 if a galaxy of the $i$th luminosity could be detected at the redshift of the $j$th
group, otherwise $N_{i,j} $ = 0.

This method, however, might suffer some problems. Not every \HATLAS
galaxy has an optical counterpart in \GAMA, even if they lie in the
same redshift range and sky region. It is possible that the dust
extinction is very large so that the galaxies are too faint in the
$r$-band to be included in \GAMA. It is also possible that a high
redshift \HATLAS galaxy is projected onto a relatively dense region in
the $r$-band selected galaxy survey so that an optical counterpart is
incorrectly assigned to it. When there are multiple optical
  candidates for a given \HATLAS galaxy, it is also possible that the
  galaxy is removed from the sample due to the optical ID being
  ambiguous. For these reasons, we also apply an alternative method
described next.

\subsubsection{Stacking Method}
Our second method (hereafter, the stacking method) is to count the
number of \HATLAS sources within a projected radius of each \GAMA
group, after subtracting the local background \HATLAS source
density. Note that for this method we do not require the \HATLAS
sources to have optical counterparts in \SDSS. Each \HATLAS source is
assigned the redshift of the target group to calculate its luminosity
and projected separation. The far-IR luminosity function in groups is
then calculated using
\begin{equation}
\Phi(L_i) \, \Delta\log L_i = 
\frac{\Sigma_{j = 0}^{N_{\rm group}}[n_{i,j}(R_j) - n_{{\rm bg},i} \times
  A_{j}]}{\Sigma_{j = 0}^{N_{\rm group}} N_{i,j}} ,
\end{equation}
where $n_{i,j}(R_j)$ is the total number of \HATLAS sources in the
$i$th luminosity bin and within a projected radius $R_j$ of the centre
of the $j$th group (the choice of $R_j$ will be discussed below),
$n_{{\rm bg},ij}$ is the background surface density of \HATLAS sources in the
$i$th luminosity bin around the $j$th group (when placed at the
redshift of the group), and $A_{j}$ is the area enclosed by radius
$R_j$.

A shortcoming of this method is that it relies on the choice of the
radius $R$ within which \HATLAS sources are counted as group
members. Ideally one would use the virial radius of the group. The
\GAMA group catalogue provides several measures of group radius,
$R_{50}$, $R_{1\sigma}$, and $R_{100}$, which are defined, respectively, 
as the radius of the 50th and 68th percentile and the most distant group
member from the central galaxy. It is possible that even the most distant
projected member is still well within the virial radius. It is beyond the
scope of this paper to investigate how well different observational
definitions of group radius reflect the ``real'' group radius. Here we
adopt $R_{100}$ as the default radius $R$. 

Another potential drawback of this method is that we need to estimate
the local background surface density $n_{\rm bg}$ of \HATLAS sources
around each group. We do this by counting sources in annuli around
each group. It is possible that the background surface density could
be over- or under-estimated if there is an over- or under-dense region
along the line-of-sight to the target group/cluster even though it is
unbiased on average. We tested how sensitive our results might be to
this effect by varying the inner and outer radii of the annuli used to
measure the local background. We find that our results are insensitive
to the exact choice of these radii (see
Appendix~\ref{sec:stacking-method}). Hereafter we use annuli of radii
$R_{100} < R < 3 R_{100}$ around each group to measure the local
background density $n_{\rm bg}$.

\section{Group luminosity lunctions}
\label{sec:LFs}
In this section, we first study the rest-frame 250$\mu$m luminosity
function for all galaxies, and then the conditional 250$\mu$m luminosity function in
galaxy groups of different masses and at different redshifts. We fit
these luminosity functions with an analytical function, and use this
to study the dependence of characteristic luminosity on group mass and
redshift, as well as the contribution to the overall 250$\mu$m
luminosity density from halos of different masses at different
redshifts.

\subsection{Far-IR luminosity function in the field}
\label{sec:field_LF}
\begin{figure}
\bc
\hspace{-0.6cm}
\includegraphics[width=8.5cm]{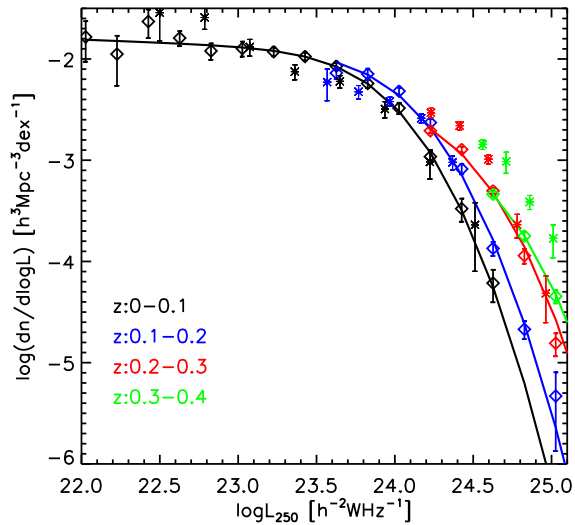}
\caption{The far-IR luminosity function of field galaxies at 250~$\mu$m. The different
  colours indicate different redshifts, as shown in the key. Diamonds
  with different colours are our results.  The curves with corresponding
  colours are the fits to our data using eqn (\ref{eq:saunders}).
  The stars show the results from \citep{Dye2010}, as dicussed in the
  text.}
\label{fig:LFfield}
\ec
\end{figure}

In order to have a better understanding of our results for the far-IR
galaxy luminosity function in different galaxy environments, we start
by measuring the 250~$\mum$ field luminosity function using all
galaxies in our matched \HATLAS-\GAMA sample with spectroscopic
redshifts. ``Field'' galaxies include galaxies in all environments. We use the $V_{\rm max}$ estimator
\citep[e.g.][]{Felten1976,Avni1980}, where the maximum volume
$V_{\rm max}$ within which a galaxy would be detected is calculated by
combining the far-IR ($S_{\nu}(250) > 35 \mJy$) and optical ($r<19.4$)
flux limits. We calculate the k-corrections for the $r-$band  using the procedure in \citet{Robotham2011}.
We calculate the luminosity function, defined as the
number of galaxies per unit volume per dex in luminosity, in 4
redshift bins over  $0<z<0.4$. Our results are shown in
Fig.~\ref{fig:LFfield} (diamonds with error bars), where different
colours show different redshifts. We estimate errorbars using the
jackknife method, dividing the full sample into 10 subsamples. We find
strong evolution in the 250~$\mum$ luminosity function even at these
low redshifts, in broad agreement with earlier work using only
\Herschel Science Demonstration Phase (SDP) data
\citep{Dye2010,Eales2010b}.

We have made a detailed comparison with the results of
\citet{Dye2010}, who used \HATLAS SDP data with a similar 250~$\mum$
flux limit, but covering only $16 \deg^2$. \citeauthor{Dye2010}'s
measurements are plotted as stars in Fig.~\ref{fig:LFfield}, for the
same redshift intervals as we use. We see that our results are in good
agreement with \citeauthor{Dye2010} for the two lowest redshift bins
at $z<0.2$, but differences start to appear in the redshift bin
$0.2<z<0.3$ and become large in our highest redshift bin $0.3<z<0.4$,
in the sense that we find weaker evolution than \citeauthor{Dye2010}
at the bright end of the luminosity function. We have identified two
main reasons for these differences: (i) \citeauthor{Dye2010}'s sample
includes \HATLAS galaxies with fainter optical counterparts ($r<22.4$)
than ours ($r<19.4$). This forces \citeauthor{Dye2010} to use less
accurate photometric redshifts for most of his sample, with
spectroscopic redshifts only for a minority of galaxies. While the
$V_{\rm max}$ method should automatically allow for the difference in
$r$-band magnitude limits between our sample and his in the case of a
uniform galaxy distribution, the strong redshift evolution of the
luminosity function breaks this assumption. Examining our highest
redshift bin, $0.3<z<0.4$, we find that our $r<19.4$ \HATLAS sample
with spectroscopic redshfit has redshifts concentrated at the lower
end of this range, while a $r<22.4$ \HATLAS sample with photo-z covers
the whole redshift bin. Due to the evolution in density across the
redshift bin, the $V_{\rm max}$ method then underestimates the mean
luminosity function in the redshift bin when we use our $r<19.4$
sample. (ii) Cosmic variance also contributes to the differences
between \citeauthor{Dye2010}'s luminosity functions and ours, since we
use the \HATLAS Phase~I catalogue, which covers a much larger area
than the SDP field used in \citeauthor{Dye2010} This allows us to
measure the luminosity function to lower far-IR luminosities in the
lowest redshift bin ($0<z<0.1$). (iii) Source completeness could also
affect the measured luminosity functions. There are three sources of
incompleteness. One is the far-IR incompleteness.  \cite{Rigby2011}
found that for the flux cut adopted in this work, 35 mJy, the
catalogue is $>$ 80\% complete. The second is the optical catalogue
incompleteness. \cite{Dunne2011} found that at $r < 21.6$ the optical
catalogue is 91.1\% complete. For our study, we use 19.4 as the
$r$-band magnitude cut, from which the completeness is even higher
than this value. The last source of incompleteness is from the
matching Herschel sources to optical galaxies. For our samples of
S$_{250\mum}>$ 35 mJy and $r<$19.4, around 80\% of the HATLAS sources
have reliable matches \citep[][and private communication]{Smith2011}
. We find that (i) dominates the differences between our LF and
\citeauthor{Dye2010}'s in the $z=0.3-0.4$ redshift bin, while effect
(ii) is the main source of differences up to $z=0.3$.  Effect (iii)
mainly matters for the faint end of the luminosity functions at
$z>0.2$.


It is convenient to describe the measured luminosity function by an
analytic fit. We use the modified Schechter function originally
proposed by \citet{Saunders1990} to fit the far-IR luminosity function
at $60 \mum$, which has a more gradual decline at high luminosity than a
Schechter function:
\begin{equation}
\phi(L) \equiv \frac{d{\rm n}}{d{\rm log}_{10}L}=
\phi^*\left ( \frac{L}{L^*}\right )^{1-\alpha}{\rm exp}[-\frac{1}{2\sigma^2} {\rm
    log}_{10}^2(1+\frac{L}{L^*})] .
\label{eq:saunders}
\end{equation}
In this function, $n$ is the number density of galaxies, $\alpha$
determines the slope at the faint end, $\sigma$ controls the shape of
the cutoff at the bright end, $L^*$ is the characteristic luminosity,
and $\phi^*$ is the characteristic density. We have fitted this function
to our measured luminosity function in each redshift bin, and the
resulting parameters are listed in Table~\ref{table:fitting}. We have
fixed the shape parameters $\alpha$ and $\sigma$ at the best-fit
values for the $z=0-0.1$ redshift bin, since our measurements at
higher redshifts do not cover a wide enough luminosity range to
robustly determine all 4 parameters in eqn (\ref{eq:saunders}). We find
that the characteristic luminosity for the $z=0-0.1$ bin is
$L^{\ast}(250) = 10^{23.67} h^{-2} \WHz$, which corresponds to a total
IR luminosity $L_{\rm IR} = 1.0 \times 10^{10} \hLsol$. Using
eqn (\ref{eq:SFR-LIR}), this corresponds to a dust-obscured SFR $=1.1
h^{-2}\Msol\yr^{-1}$. Based on our fits, $L^{\ast}(250)$ increases
rapidly with redshift, being about 3 times larger at $z=0.35$ compared
to $z=0.05$. The characteristic density $\phi^{\ast}$ also changes
rapidly with redshift, falling by a factor 7 over the same redshift
range.

\begin{table*}
  \caption{Best fitting luminosity function parameters for the 250$\mum$
    field LF at different redshifts, using eqn (\ref{eq:saunders}). $\alpha$ and $\sigma$ are
    fixed using the fit at $z=0-0.1$.}

\begin{tabular}{||l||c||c||c||c||}
\hline
redshift   & $\alpha$ & $\sigma$   & ${\rm log} \phi^*$[$h^3$Mpc$^{-3}$dex$^{-1}$] & log L$^*$[$h^{-2}$WHz$^{-1}$] \\
\hline       
0  - 0.1        & 1.06  & 0.30 & -1.91$\pm$0.04     &  23.70$\pm$0.07      \\
0.1 - 0.2      & 1.06  & 0.30 & -1.94$\pm$0.06   &  23.83$\pm$0.04   \\
0.2 - 0.3      & 1.06  & 0.30 & -2.39$\pm$0.05     & 24.14$\pm$0.03   \\ 
0.3 - 0.4      & 1.06  & 0.30  & -2.72$\pm$0.04   &  24.30$\pm$0.03    \\
\hline
\end{tabular}
\label{table:fitting}
\end{table*}

\subsection{Far-IR luminosity function in groups}

\begin{figure*}
\bc
\includegraphics[width=18cm, bb= 28 40 600 622]{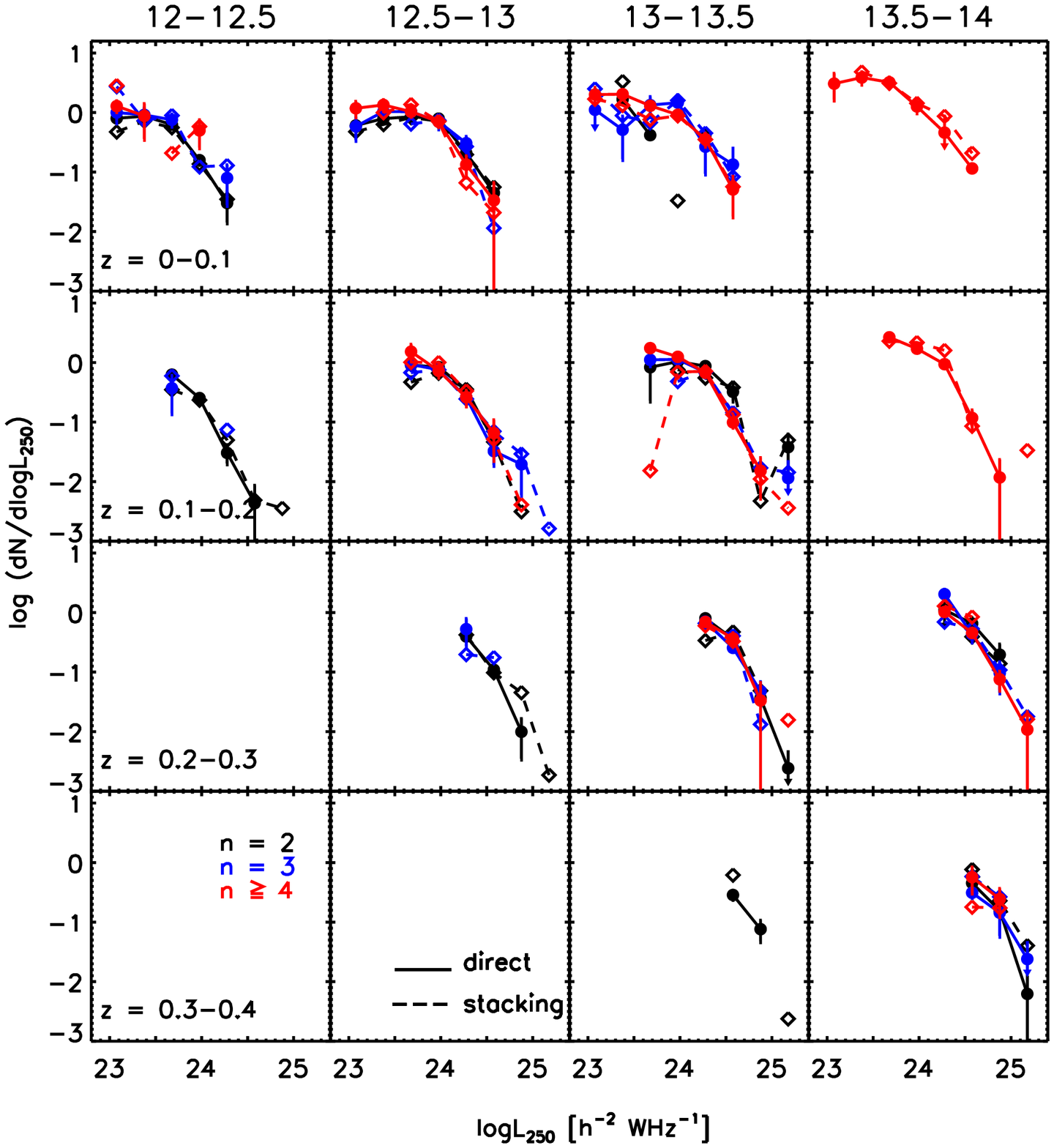}
\caption{Group far-IR conditional luminosity function (defined as mean
  number of galaxies per group) at 250~$\mu$m as a function of halo
  mass (left to right) and redshift (top to bottom). The redshift
  range is indicated in the left panels, and the logarithm of the
  group halo mass in $\hMsol$ above the top panels. Solid curves with
  filled circles show results from the direct method, and dashed
  curves with empty diamonds from the stacking method. For clarity,
  errorbars are plotted only for the direct method. The different
  colours are for groups of different optical multiplicities, $n$: black
  for multiplicity $n=$2, blue for $n=$3 and red for $n\geq
  4$. Errorbars which extend down to 0 are indicated with downward arrows.}
\label{fig:clf_comparison}
\ec
\end{figure*}

We begin our analysis of the far-IR conditional luminosity function
(CLF) in galaxy groups by comparing results obtained using the two
methods described in \S\ref{sec:CLF_methods}, the {\em direct method}
and the {\em stacking method}. We split our sample according to group
mass and redshift, in order to separate environmental effects from
redshift evolution. The results are shown in
Fig. \ref{fig:clf_comparison}, with the direct method shown by solid
lines and the stacking method by dashed lines. We have estimated
errorbars using the jackknife method. We have also checked for any
dependence of the CLFs on the group optical multiplicity $n$, defined
as the number of $r$-band selected galaxies with spectroscopic
redshifts which define this group in the \GAMA group catalogue. The
multiplicity $n$ is relevant both for the estimate of the group radius
(which is important for the stacking method) and for the estimate of
the group mass. We thus further split our group samples according to
their multiplicities. Results for different multiplicities are shown
by different colours (black: $n=2$, blue: $n=3$, and red: $n\geq 4$).
For groups of a given mass, redshift range, and multiplicity,
Fig.~\ref{fig:clf_comparison} shows that the luminosity functions
estimated using the direct and stacking methods are consistent.
  Since the results from the {\em stacking method} are independent of
  the optical ID matching used in the {\em direct method}, the consistency
  between the CLFs obtained by the two methods demonstrates that the
  {\em direct method} does not miss a significant fraction of \HATLAS
  galaxies due to there being either multiple optical candidates for
  a given \HATLAS galaxy, or no counterparts brighter than
  $r<19.4$. However, the results from the stacking method are much
noisier, which appears to be due mostly to uncertainties in the
background subtraction. We therefore use only results from the direct
method in the analysis that follows. We also find that for both
methods, the inferred group luminosity functions depend only weakly on
the optical multiplicities. In the following analysis we therefore use
all groups (i.e. multiplicity $n \geq 2$) in order to have better
statistics, unless indicated otherwise. We see that conditional
luminosity functions can be measured over quite a wide range of group
halo mass ($10^{12}-10^{14} \hMsol$) for $z<0.2$, though this range
shrinks with redshift, so that in the highest redshift bin we can
measure luminosity functions only in the most massive groups.

\subsection{Characteristic properties of the far-IR luminosity
  function in groups}

\begin{figure*}
\bc
\hspace{-0.6cm}
\includegraphics[width=18cm, bb= 28 40 600 622]{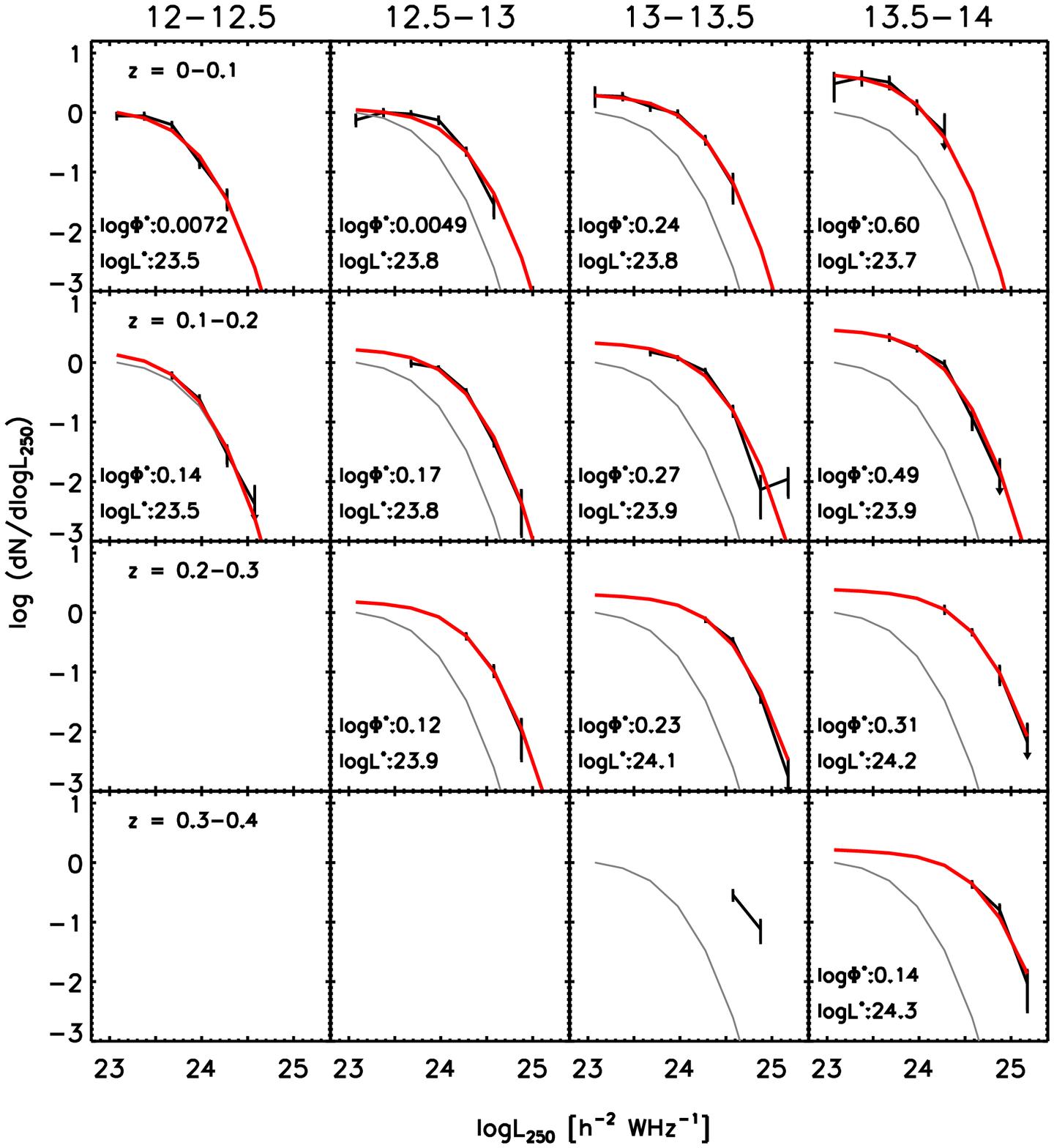}
\caption{Far-IR luminosity functions of galaxies in groups. Black
  curves with error-bars are the direct measurement for optical
  multiplicity $n\geq 2$, and red curves are the analytic fits. The
  parameters $\alpha$ and $\sigma$ in eqn (\ref{eq:saunders}) are fixed
  to 1.06 and 0.30 as measured for the field luminosity function at $z
  = 0 - 0.1$. The other two parameters encoding the characteristic
  luminosity $L^{\ast}$ (in units of $h^{-2} \WHz$) and the amplitude
  of the luminosity functions $\Phi^{\ast}$ are indicated in the lower
  left corner of each panel.  The redshift range is indicated in the
  left panels, and the logarithm of the group halo mass in $\hMsol$
  above the top panels.  The fit in the top left panel is replicated
  as a grey curve in all the other panels as a reference.  }
\label{fig:clf_fitting}
\ec
\end{figure*}

In order to study the dependence of the far-IR luminosity function on
group mass and redshift, it is convenient to fit the measured
conditional luminosity functions with an analytical function. We use
the same modified Schechter function as in eqn (\ref{eq:saunders}),
except with $\Phi$, the mean number of galaxies per group per
$\log_{10}L$, replacing $\phi$, the mean number of galaxies per unit
volume per $\log_{10}L$, and correspondingly $\Phi^{\ast}$ replacing
$\phi^{\ast}$. Since our measured CLFs mostly do not cover a wide
enough range in luminosity to reliably constrain all 4 parameters
($\alpha$, $\sigma$, $\Phi^{\ast}$ and $L^{\ast}$) in the fit, we fix
the shape parameters at the values $\alpha=1.06$ and $\sigma=0.30$
which we measure for the $z=0-0.1$ field LF, and then fit
$\Phi^{\ast}$ and $L^{\ast}$ independently for each bin in group mass
and redshift. The resulting fits are shown in
Fig. \ref{fig:clf_fitting}, where the black curves with error bars show
the direct measurements, while the red curves show the fits.  The
measured CLF for the mass range of $10^{13} - 10^{13.5}$ and redshift
range $z = 0.3 - 0.4$ has only two data points, so we do not try to fit
this with our analytic function. It can be seen that the functional
form of eqn (\ref{eq:saunders}) provides a good fit to our measured
CLFs for all mass and redshift ranges for which we have data. To show
the dependence of the CLF on group mass and redshift more clearly, we
also repeat the fit from the top left panel ($z=0-0.1$ and $M_{\rm h} =
10^{12} - 10^{12.5} \hMsol$) as a grey line in the other panels. This
shows that the CLF tends to increase with both group mass and
redshift.  The far-IR
multiplicity, measured by the number of galaxies with $L_{250} >
10^{23.5} h^{-2} \WHz$ is around unity in the least massive groups in our
sample.

\begin{figure}
\bc
\hspace{-0.6cm}
\includegraphics[width=8cm]{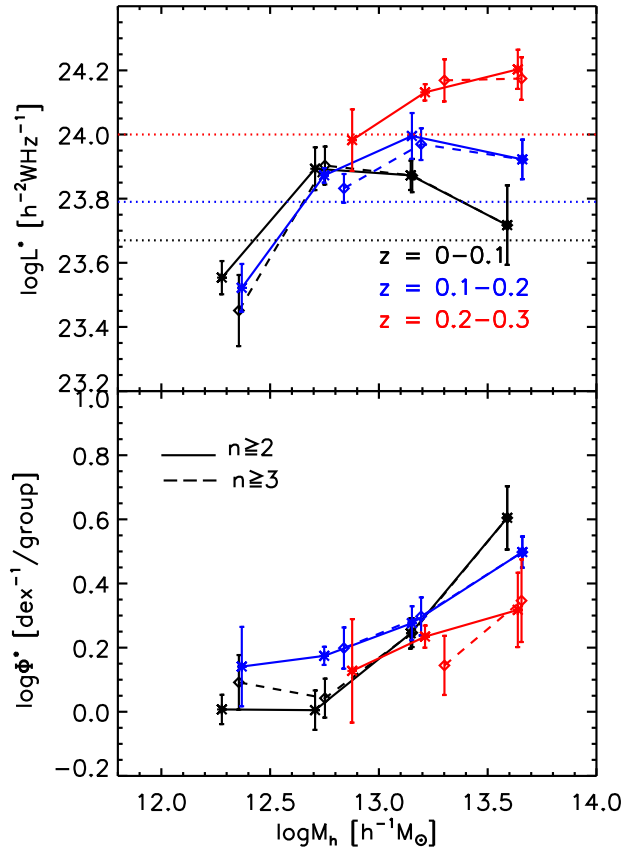}
\caption{Top panel: characteristic luminosity, $L^{\ast}$, as a
  function of group mass and redshift. Different colours show different
  redshift bins, as indicated by the legend. The crosses and solid
  lines show results for optical multiplicity $n \geq 2$, and open
  diamonds and dashed lines for $n \geq 3$. The error bars show
  jackknife errors. the dotted horizontal lines show $L^{\ast}$ for
  the field luminosity function at the same redshifts. Bottom panel:
  normalization $\Phi^{\ast}$ as a function of group mass and
  redshift. In both panels, data are plotted at the median value of
  group mass for that bin. }
\label{fig:Lstar-phistar}
\ec
\end{figure}

We show the dependence of the CLF parameters $L^{\ast}$ and
$\Phi^{\ast}$ on group mass and redshift in the top and bottom panels
of Fig. \ref{fig:Lstar-phistar}. (We omit the results for the redshift
bin $z=0.3-0.4$ from this and the following plots, since we have only
measured the CLF for two bins in group mass for this case.) The
solid lines show results for group optical multiplicity $n \geq 2$ and
the dashed lines for $n \geq 3$. We see that the results for different
multiplicity cuts are generally consistent for both $L^{\ast}$ and
$\Phi^{\ast}$.

Examining first the dependence of $L^{\ast}$ on group mass, we see that for $z<0.2$,
it increases steeply with group mass at low masses, but then appears
to turn over to a gradual decline at high masses, although the large
errorbars on $L^{\ast}$ for high masses make it difficult to be
certain about the decline. Note that the estimated completeness of the group catalogue for
the lowest mass range is rather low, $\sim$ 45\% and 20\% at 0$<z<$0.1 and 0.1$<z<$0.2, respectively. 
This might lead to an overestimation of the dependence on group mass at these masses. 
For $0.2<z<0.3$, only groups more massive than $\sim
10^{12.5} \hMsol$ are detected in \HATLAS-\GAMA. For this redshift
range, the measured $L^{\ast}$ increases monotonically with group
mass, though appearing to flatten at the highest masses.

The redshift evolution of $L^{\ast}$ thus depends strongly on the
group mass. For the highest masses sampled, $M_{\rm h} \sim 10^{13.75}\hMsol$ 
(i.e. clusters), $L^{\ast}$ increases by a factor 2-3 over the
range $0.05 < z < 0.35$, while for more typical groups, with 
$M_{\rm h} \sim 10^{12.75} \hMsol$, there is almost no evolution for $0.05 < z < 0.25$.

In the top panel of Fig. \ref{fig:Lstar-phistar}, we also overplot as
horizontal dotted lines the values of $L^{\ast}$ which we measure for
the field LF at the same redshifts. We see that $L^{\ast}$ for field
galaxies always lies between the values in the least and the most
massive groups, consistent with the finding in previous work that most
\HATLAS galaxies resides in groups of mass comparable to the Milky
Way halo \citep{Guo2011b}. 

In the bottom panel of Fig. \ref{fig:Lstar-phistar} we show how the CLF
normalization $\Phi^{\ast}$ varies as a function of group mass at
different redshifts. We again see that the redshift evolution depends
on group mass. For lower mass groups ($M_{\rm h} \sim 10^{12.25} -
10^{12.75} \hMsol$), $\Phi^{\ast}$ increases with redshift for
$z<0.2$, while for the highest masses ($M_{\rm h} \sim 10^{13.75} \hMsol$,
it appears instead to decrease with increasing redshift for $z \lsim
0.3$, although the large errorbars in the latter case make it
difficult to be certain about the behaviour.

\subsection{Far-IR luminosity-to-mass ratio of groups, and the far-IR
  luminosity density}

A further important physical quantity which we can calculate from our
measured group far-IR CLFs is the total far-IR luminosity-to-mass
ratio of groups, since this is related to the dust-obscured SFR per
unit dark halo mass. Previous studies have found that the fraction of
star-forming galaxies decreases with group mass
\citep[e.g.][]{Dressler1980,Kimm2009}. However, direct measurements of
SFR per group mass are very rare because the determination of the SFR
depends greatly on corrections for dust extinction when using UV and optical tracers,
and also because it is not trivial to measure group masses for large
samples.

\begin{figure*}
\bc
\hspace{-0.6cm}
\includegraphics[width=16cm]{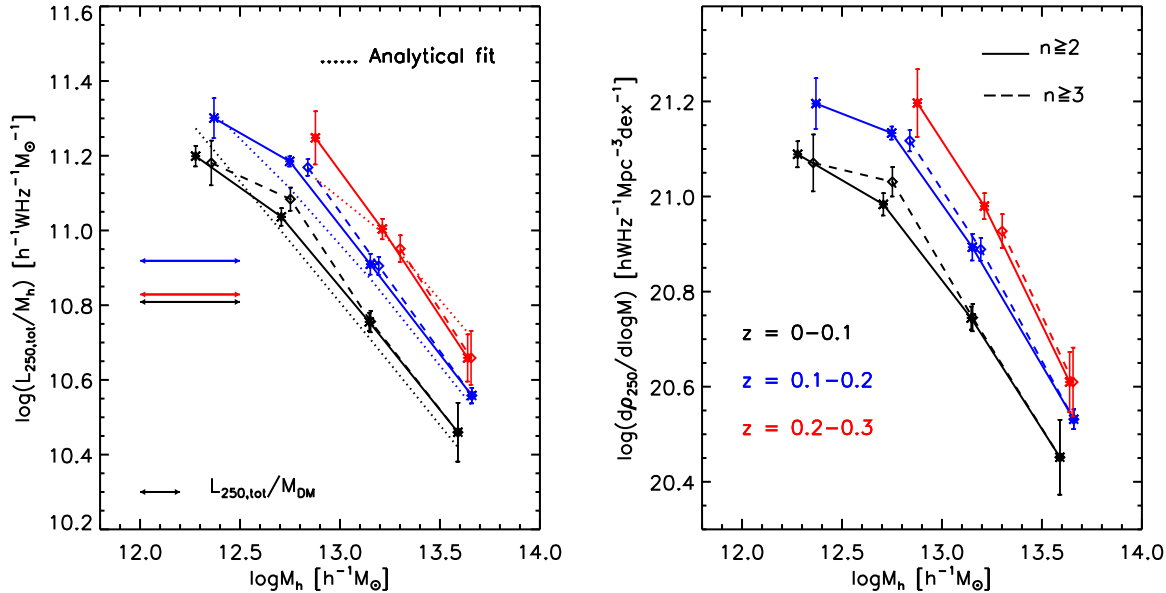}
\caption{Left panel : The ratio of the total far-IR luminosity at
  250~$\mum$ to total group mass as a function of group mass and
  redshift.  Right panel: Contribution to the far-IR (250~$\mum$)
  luminosity density from halos of different mass. In both panels,
  different colours are for different redshift ranges, errorbars are
  estimated using the jackknife technique, and solid and dashed lines
  are for group optical multiplicity $n \geq 2$ and $n \geq 3$,
  respectively.  The dotted lines in the left panel show the analytic
  fit, eqn (\ref{eq:L250toM_fit}), evaluated at the median mass and
  redshift for each bin. In both panels, data are plotted at the
  median value of group mass for that bin in $\log M_{\rm h}$ and
  redshift. The horizontal lines in the left panel show the ratio of
  250~$\mum$ luminosity density to dark matter density for the whole
  galaxy population, calculated from the field 250~$\mum$ LF. }
\label{fig:LtoM}
\ec
\end{figure*}

Here we integrate our analytic fits to the group CLFs shown in
Fig. \ref{fig:clf_fitting} over luminosity to estimate the average
total 250~$\mum$ luminosity $L_{250,\rm tot}$ for groups in each mass and
redshift range, and hence obtain the 250~$\mum$ luminosity-to-mass
ratios of groups. Since we have not directly measured the group CLFs
at $L_{250} < 10^{23} h^{-2} \WHz$, but instead simply assumed the
same faint-end slope $\alpha$ as we measured for the field 250~$\mum$
LF at $z<0.1$, we calculate the total group luminosities $L_{250,\rm tot}$
using two different lower limits of integration, $L_{250, \rm min} =0$ and
$10^{23.5} h^{-2} \WHz$. The values of total luminosity drop by a
factor of up to 1.5 when using the higher luminosity cut. Our results
for the luminosity-to-mass ratios $L_{250,\rm tot}/M_{\rm h}$ for $L_{250, \rm min} =0$
are shown in the left panel of Fig. \ref{fig:LtoM}.  As
in Fig.~\ref{fig:Lstar-phistar}, the solid lines show results for
group optical multiplicity $n \geq 2$, and dashed lines are for $n
\geq 3$, from which we see that our estimates of $L_{250,\rm tot}/M_{\rm h}$ are
insensitive to optical multiplicity. We also note that there is some
degeneracy between our fitted values of $L^{\ast}$ and $\Phi^{\ast}$
in the group CLFs, but the effects of this are partly removed when we
calculate the luminosity-to-mass ratios, which is reflected in the
size of the errorbars plotted in Fig. \ref{fig:LtoM}.

Fig. \ref{fig:LtoM} shows that at each redshift, the 250~$\mum$
luminosity-to-mass ratio is a decreasing function of the group
mass. At $z = 0$, the maximum $L_{250,\rm tot}/M_{\rm h}$ is $\sim 10^{11.2}
h^{-2} \WHz$ in groups with masses $\sim 10^{12.3} \hMsol$, comparable
to the Milky Way halo, and decreases to $\sim 10^{10.5} h^{-2} \WHz$
for groups of mass $\sim 10^{13.5} \hMsol$. This implies a decreasing
rate per unit mass for converting baryons to stars through
dust-obscured star formation with increasing group mass.  The
dependence of $L_{250,\rm tot}/M_{\rm h}$ on group mass can be fitted by a
power-law except at the very low mass end, where the slope becomes
flatter. At higher redshift, the $L_{250,\rm tot}/M_{\rm h}$ vs $M_{\rm h}$ relation
shares the same slope as that at $z= 0$, while its amplitude increases
significantly with redshift. Specifically, the amplitude increases by
about a factor of 3 from $z \approx 0.05$ to $z \approx 0.35$. We fit this
luminosity-to-mass ratio as a function of group mass and redshift with
the following equation:

\begin{equation}
\frac{L_{250, {\rm tot}}}{M_{\rm h}} = 
10^{11.3\pm0.3}  (1+z)^{4.74\pm0.41} (\frac{M_{\rm h}}{10^{12} \hMsol})^{-0.65\pm0.02}.
\label{eq:L250toM_fit}
\end{equation}

This analytic fit is shown by dotted lines in the left panel of
Fig. \ref{fig:LtoM}, where it is evaluated and plotted for the median
mass and redshift of the groups in each bin. While the
luminosity-to-mass ratio $L_{250,\rm tot}/M_{\rm h}$ decreases with group mass,
the far-IR luminosity increases with mass roughly as $L_{250,\rm tot}
\propto M_{\rm h}^{0.35}$ over the range of mass $10^{12} < M_{\rm h} < 10^{14}
\hMsol$ probed in this study. 

The horizontal lines in the left panel of Fig. \ref{fig:LtoM} show the
mean luminosity-to-mass ratio $L_{250,\rm tot}/M_{\rm DM}$ for the galaxy
population as a whole at the same redshifts, obtained by integrating
over the field LF and dividing by the cosmological dark matter
density. We see that $L_{250,\rm tot}/M_{DM}$ in the field increases with
redshift in a similar way to that in groups between the two lowest
redshift bins, but then drops in the $z=0.2-0.3$ bin. This drop may be
caused by errors in our estimate of the field LF in this redshift
range, as discussed in \S\ref{sec:field_LF}.

Finally, we combine our measurement of the far-IR luminosity-to-mass
ratios of groups with a theoretical prediction for the number density
of halos as a function of mass to estimate the contribution to the
far-IR luminosity density at 250~$\mum$, $\rho_{250}$, from groups of
different masses:
\begin{equation}
\frac{d\rho_{250}}{d\log M_{\rm h}} =  \left( \frac{L_{250,\rm tot}}{M_{\rm h}} \right)
M_{\rm h} \left( \frac{dn}{d\log M_{\rm h}} \right)
\label{eq:rho250}
\end{equation}
In the above formula, we use the theoretically predicted dark matter
halo mass function $dn/d\log M_{\rm h}$ in a standard $\Lambda$CDM
cosmology, specifically, the analytical mass function of
\citet{Reed2007}, which has been shown to match N-body simulations
very well. We also use the directly measured values of
$L_{250,\rm tot}/M_{\rm h}$ for each bin in mass and redshift, rather than the
analytical fit in eqn (\ref{eq:L250toM_fit}).

The right panel of Fig.~\ref{fig:LtoM} shows the resulting estimate of
the contribution to the 250~$\mum$ luminosity density $\rho_{250}$
from groups of different mass. As in the left panel, results are split
into different redshift bins. We find that the total far-IR luminosity
density contributed by halos of different masses is a decreasing
function of halo mass. For those more massive than $10^{12.5} \hMsol$,
the far-IR luminosity density can be fitted with a power law, while the slope gets flatter for
lower masses at $z<0.2$, where we still have measurements for such low
masses. The 250~$\mum$ luminosity density increases with redshift at
all group masses. This
behaviour is very similar to that of the luminosity-to-mass ratio,
which is expected since the evolution of the dark matter halo mass
function in this redshift range is quite weak. As before, we find that
our results are insensitive to whether we use groups with optical
multiplicity $n \geq 2$ or $n \geq 3$.

For completeness, we also derive the total far-IR luminosity density
$\rho_{250}$ by integrating over our measured field galaxy luminosity
function, and report our results in Table~\ref{table:rho250}. We also
report there our estimates of the contributions to the total
luminosity density from groups in the mass ranges probed by the
\HATLAS-\GAMA survey. We give values of $\rho_{250}$ for two different
lower limits for the integrations over $L_{250}$, $L_{250, \rm min} =0$ and
$10^{23.5} h^{-2} \WHz$. We find that whichever of these luminosity
cuts we adopt, groups more massive than $10^{12}\hMsol$ contribute
around 70\% of the total luminosity density at $z<0.2$. For
$0.2<z<0.3$, groups more massive than $10^{12.5} \hMsol$ already
contribute nearly 70\% of the total.  

\begin{table*}
\caption{Integrated 250$\mum$ luminosity density from the field and
  from groups as a function of redshift. The first column is the
  redshift range. The second and the third columns give the total
  luminosity density in the field from integrating our analytic fit
  down to $L_{250, \rm min} = 0$ and $10^{23.5} h^{-2} \WHz$,
  respectively. The fourth column gives the ranges of group halo mass
  probed in our study. The fifth and sixth columns give the
  contributions to the total luminosity density from groups in these
  mass ranges, using the same two lower luminosity cuts as for the
  field. The percentages give the fractions of the corresponding field
  luminosity density for the same lower luminosity cut. Halo masses
  are given in units $\hMsol$ and luminosity densities in units $h
  \WHz\Mpc^{-3}$.  }

\begin{tabular}{||l||c||c||l||l||l||}
\hline
redshift   & Field & Field ($>10^{23.5}$)   & log M$_{\rm h}$ & Group & Group ($>10^{23.5}$) \\
\hline       
0  - 0.1        & 10$^{21.65}$  & 10$^{21.45}$ & $ >$ 12      &  $10^{21.48}$ (68\%)   &   $10^{21.30}$ (70\%)   \\
0.1 - 0.2      & 10$^{21.75}$  & 10$^{21.60}$ & $ >$ 12   &  $10^{21.60}$ (71\%)   &   $10^{21.43}$ (67\%)  \\
0.2 - 0.3      & 10$^{21.72}$  & 10$^{21.64}$ & $ >$ 12.5     &  $10^{21.46}$ (55\%)   &   $10^{21.38}$ (55\%) \\ 
\hline
\end{tabular}
\label{table:rho250}
\end{table*}


\subsection{Comparison with previous work} 


Previous direct measurements of the IR LFs of galaxy groups and
clusters are quite limited.  \cite{Bai2006,Bai2007,Bai2009} used
mid-IR (Spitzer 24$\mum$) data to measure the IR LFs of several rich
clusters ($M \sim 10^{15} \Msol$) at $z \lsim 1$ and found strong
redshift evolution in $L_{\ast}$, as also found in the field
\citep[e.g.][]{LeFloch2005}, but no dependence on radius within a
cluster. They also found that the shape of the IR LF in clusters was
similar to that in the field at the same redshift, a result confirmed
by \cite{Finn2010}, who studied 16 clusters drawn from the ESO Distant
Cluster Survey at 0.4$<z<$0.8.  This is however, in constrast to what was 
found by \cite{Goto2010} who used the \AKARI\ 8$\mum$ observations of a
single rich cluster at $z \sim 0.8$, and found that $L_{\ast}$ is
lower by a factor 2.4 compared to the field at the same
redshift. Comparing with our results,  Fig. \ref{fig:Lstar-phistar}
shows that in clusters with $M_{\rm h} \sim 10^{14} {\rm \Msol}$, $L_{\ast}$
differs by less than 50\% from the field value, while the difference
can be larger at lower group masses.

H$\alpha$ is another important tracer of the SFR. The H$\alpha$ LFs of
rich clusters have been measured in various studies, and generally
been found to have similar shapes to that of the field population at the
same redshift \citep[e.g.][]{Balogh2002,Kodama2004}. This is similar
to the result for IR LFs. In a related result,
\cite{Giodini2012} measured the stellar mass function of
star-forming galaxies in galaxy groups with $10^{13} \lsim M \lsim
10^{14} {\rm \Msol}$ at $0.2<z<1$, and found that it has a similar shape to
that for the field.

As discussed above, $L_{\rm IR,\rm tot}/M_{\rm h}$ is an indicator of the
total dust-obscured SFR (summed over all galaxies) per unit halo
mass. For our sample of $\sim 3000$ galaxy groups with $10^{12} < M_{\rm h}
< 10^{14}  \hMsol$ at $z<0.4$, we find $L_{\rm IR,\rm tot}/M_{\rm h}
\propto (1+z)^5$ (see Fig. \ref{fig:LtoM}). This dependence is in
reasonable agreement with that found in previous work from mid-IR
observations of samples of clusters ($M_{\rm h} \gsim 10^{14} {\rm \Msol}$) for
$0<z<1$, which found $\Sigma SFR/M_{\rm h} \propto (1+z)^{\alpha}$,
with $\alpha \approx 5-7$ \citep{Geach2006, Bai2007, Bai2009,
Koyama2010, Webb2013}. Based on data from  the Herschel Multi-tiered 
Extra-galactic survey \citep{Oliver2010}, \cite{Bernardis2012} also find  
a similar redshift dependence, $\alpha \sim 4$ for z = 0.2 - 4, using a HOD (Halo Occupation Distribution model) fitting method. 
Our result is also similar to that found by
\citet{Popesso2012} from \Herschel far-IR (100 and 160$\mum$)
observations of a sample of $\sim 20$ massive groups and rich clusters
($10^{13} \lsim M_{\rm h} \lsim 10^{15} \Msol$) at $0.1 \lsim z \lsim 1$. We
note that these previous studies all estimated $\Sigma SFR$ by summing
IR-based SFRs over galaxies brighter than some IR luminosity limit,
typically $L_{\rm IR} \gsim 10^{11} \Lsol$. In contrast, we fit the IR LFs
of groups down to much fainter luminosities, and then integrate over
these fits (extrapolated to $L_{\rm IR}=0$) to estimate the total group IR
luminosities. Since the characteristic IR luminosity $L_{\ast}$, and
hence the shape of the LF, evolves with redshift, these two approaches
will lead to redshift evolution factors that differ in
detail. Indications of similarly strong evolution of $\Sigma
SFR/M_{\rm h}$ were also found from studies using H$\alpha$-based
SFRs, for small samples of clusters at $0.2 \lsim z \lsim 0.8$
\citep[e.g.][]{Kodama2004,Finn2004, Finn2005}.

From our sample of galaxy groups, we also find a dependence
on group mass, $L_{\rm IR,\rm tot}/M_{\rm h} \propto M_{\rm h}^{-0.65}$. This
trend is qualitatively similar to indications from previous H$\alpha$
\citep{Finn2005}, mid-IR \citep{Bai2007} and far-IR \citep{Popesso2012}
studies of galaxy clusters, although it is significantly flatter than
the trend $\Sigma SFR/M_{\rm h} \propto M^{-1.5\pm0.4}$ found by
\citet{Webb2013} from mid-IR observations of a sample of clusters at
$0.3<z<1$. (Note, however, that the \citeauthor{Webb2013} estimates of
$\Sigma SFR/M_{\rm h}$ only include galaxies brighter than $L_{\rm IR} >
2\times 10^{11} \Lsol$.) Compared to previous work, our study,
although restricted to a lower redshift range, covers a much lower and
wider range of group mass and a wider range of IR luminosity, as well
as having much better statistics due to the larger number of groups.

\section{Far-IR - optical colours  in groups}
\label{sec:colours}
The far-IR emission is a good indicator for the dust-obscured SFR,
since it represents the energy re-emitted by dust when heated by
(mostly young) stars. On the other hand, the optical luminosity is a
tracer of the stellar mass, since it includes emission from older
stars. The 250~$\mum$ to $r$-band colour should therefore be a good
indicator of the specific star formation rate ($sSFR =SFR/M_{\star}$). The dividing line between ``star-forming'' and
``passive'' galaxies is typically defined as $sSFR>{10^{-11}
\yr^{-1}}$ \citep[e.g.][]{Weinmann2010}. At the median redshift $z \approx 0.2$ of the matched
\HATLAS-\GAMA sample, the flux limit $S_{\nu}(250\mum) > 35$mJy of
our far-IR-selected sample corresponds to a dust obscured SFR $\sim 4
{\rm \Msol}\yr^{-1}$. Therefore, galaxies included in our \HATLAS-\GAMA
sample would typically be classed as star-forming based on their sSFR,
provided they have stellar masses $\lsim 10^{11} \Msol$.

\begin{figure*}
\bc
\hspace{-0.6cm}
\resizebox{12cm}{!}{\includegraphics{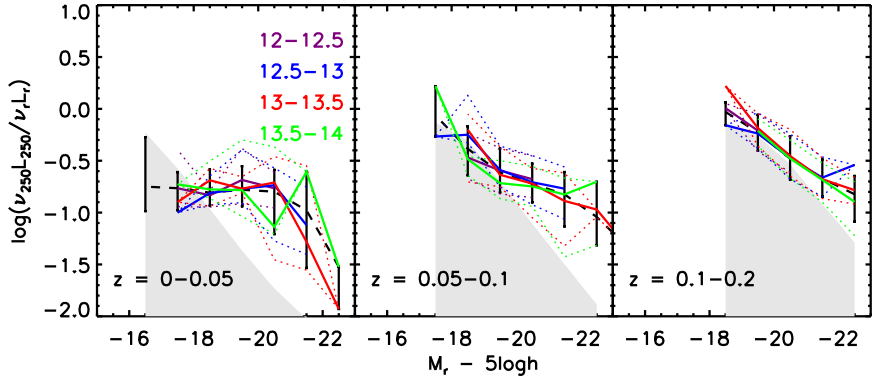}}
\caption{The far-IR-to-optical colour vs. $r$-band absolute magnitude
  relation for different environments. The left, middle and right
  panels are for redshift ranges $0<z<0.05$, $0.05<z<0.1$ and
  $0.1<z<0.2$ respectively as labelled. The black dashed lines show the
  colour-magnitude relation for field galaxies in that redshift
  range. The coloured lines show the relation for galaxies in groups
  of different masses, with the logarithm of the group mass (in
  $\hMsol$) being given by the key in the left panel. The thick lines
  show the median colour, and the thin dotted lines indicate the 68\%
  range around the median. Grey regions indicate the region within
  which incompleteness due to the 250~$\mum$ flux limit is
  important. }
\label{fig:color}
\ec
\end{figure*}

We use galaxies from our matched \HATLAS-\GAMA sample, to obtain both
far-IR and the optical luminosities, and hence their far-IR-to-optical
colours. We further restrict our analysis to $z<0.2$. In
Fig.~\ref{fig:color}, we plot the rest-frame 250~$\mum$-to-$r$-band
luminosity ratio $(\nu_{250}L_{250})/(\nu_r L_r)$, which is an indicator
of sSFR, against the $r$-band absolute magnitude, which is an
indicator of stellar mass. The three panels show different redshift
ranges. In each panel, the dashed black lines show the median
250~$\mum$/$r$-band colour for all \HATLAS-\GAMA galaxies in that bin
of $r$-band absolute magnitude, with the errorbars showing the 16-84\%
range around this (equivalent to the 1$\sigma$ range for a
Gaussian). The coloured lines show the median colours for galaxies in
groups of different masses, as indicated by the key, with the dotted
lines indicating the 16-84\% range. The grey region in each plot
indicates where our \HATLAS-\GAMA sample becomes significantly
incomplete due to the 250~$\mum$ flux limit. We calculate the upper
boundary of this region in each bin of absolute magnitude $M_r$ from
the 250~$\mum$ luminosity for a galaxy at the 250~$\mum$ flux limit at
the median redshift for all \GAMA galaxies in that absolute magnitude
bin in that redshift range (whether they are detected at 250~$\mum$ or
not), assuming a median dust temperature of 26~K. This provides only
 a rough estimate of the completeness
boundary, since some galaxies will be at redshifts lower than the
median, and so would be detected with lower 250~$\mum$ luminosities than the
simple estimate above, and bacause the 1 $\sigma$ scatter in galaxy temperature
 could be as large as 4~K. This effect explains why the median
colour-magnitude relation for faint $M_r$ falls just inside the grey
incompleteness region in the $0<z<0.05$ panel - in these cases, the
median redshift of the matched \HATLAS-\GAMA sample is below the
median redshift for the full \GAMA sample at the same $M_r$.

We see from the left panel of Fig.~\ref{fig:color} that in the lowest
redshift range, $0<z<0.05$, the median colour vs. magnitude relation
does not depend on group mass over the whole mass range $10^{12} < M_{\rm h}
< 10^{14} \hMsol$, and is indistinguishable from the relation for all
galaxies. The median far-IR-to-optical colour also depends only weakly
on $r$-band absolute magnitude. The scatter around the
colour-magnitude relation in groups also appears to be very similar to
that for the field.


For the highest redshift range $0.1<z<0.2$, shown in the right panel
of Fig.~\ref{fig:color}, our estimated colour-magnitude relation and
scatter lie just above the completeness boundary at all absolute
magnitudes. We conclude from this that our measured colour-magnitude
relation in this redshift range is probably determined mostly by the
250~$\mum$ flux limit of the \HATLAS survey. Therefore we cannot draw
any firm conclusions about the real form of the colour-magnitude
relation or its dependence on group mass at redshift $z>0.1$ from
these data. For the intermediate redshift range $0.05<z<0.1$, shown in
the middle panel of Fig.~\ref{fig:color}, our median colour-magnitude
relation falls on the selection boundary at faint magnitudes, and the
lower 10-percentile value is close to the selection boundary even for
brighter magnitudes. It therefore seems likely that the tilt in our
estimated colour-magnitude relation in this redshift range is also
mainly due to selection effects in the sample.

\begin{figure*}
\bc
\hspace{-0.6cm}
\resizebox{12cm}{!}{\includegraphics{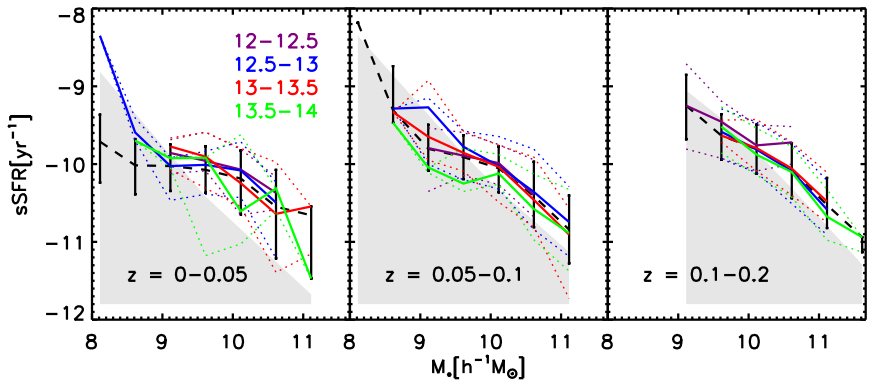}}
\caption{The specific star formation rate $SFR/M_{\rm star}$ vs stellar
  mass $M_{\rm star}$. The line type and colour coding are the same as in Fig.~\ref{fig:color}.
}
\label{fig:SSFR}
\ec
\end{figure*}

We further convert the 250 $\mu$m luminosity to SFR according to 
Eq.(\ref{eq:SFR-LIR}). The galaxy stellar mass is calculated using the g-i 
colour and i-band luminosity, following the procedure in \cite{Taylor2011}, and the SFR 
is calculated using individual temperatures for each source from SED fitting. We assume the Chabrier IMF for both the SFR and the stellar mass.  The 
corresponding sSFR vs. stellar mass relations at different redshifts are
 presented in Fig.~\ref{fig:SSFR}. As in Fig.~\ref{fig:color}, it shows in the lowest 
redshift range , $0<z<0.05$,  the median values of the sSFR as a function of galaxy stellar
mass are indistinguishable between halos of different masses, and the difference from those
for all galaxies is very small.  The scatter around the sSFR vs. stellar mass relation is 
similar in groups of different mass, and also similar to those for the field. These results suggests
 that the sSFR vs stellar mass relation for dust-obscured star formation is almost independent
of host halo mass for group masses $M_{\rm h} < 10^{14} \hMsol$. Results for higher
redshifts are limited by the selection effect (gray region) as in Fig.~\ref{fig:color} and thus no firm 
conclusions could be drawn from current data.

Our result is therefore that the far-IR/optical colour, and the sSFR for dust-obscured star formation, are
independent of the group mass at a given optical luminosity or stellar
mass. This is consistent with most previous work on the dependence of
sSFR for star-forming galaxies on environment in the local universe,
using a variety of star formation tracers and measures of galaxy
environment. Early studies using the H$\alpha$ equivalent width
(EW) as an indicator of sSFR found that this is independent of local
galaxy density for the star-forming population, even though the
fraction of galaxies classed as star-forming does change with
environment \citep{Balogh2004,Tanaka2004}. \citet{Weinmann2006} used
emission line SFRs to show that the sSFRs of actively star-forming
galaxies at a given stellar mass depend only weakly on host halo mass
over the range $10^{12} \lsim M_{\rm h} \lsim 10^{15} \Msol$, and
\citet{Peng2010} found a weak dependence on local galaxy density using
similar data.  \citet{Bai2010} and \citet{McGee2011} used SFRs based
on mid-IR and far-UV data respectively to show that the sSFRs of
star-forming galaxies in groups were similar to those of field
galaxies, although \citeauthor{Bai2010} also found lower sSFRs in rich
clusters. Indications of lower average SFRs for star-forming galaxies
in clusters have also been found in some H$\alpha$ studies
\citep{Gomez2003,Finn2005}. 



\section{Comparison with Galaxy Formation Model Predictions}
\label{sec:models}

\begin{figure*}
\bc
\hspace{-0.6cm}
\includegraphics[width=18cm, bb= 28 40 600 622]{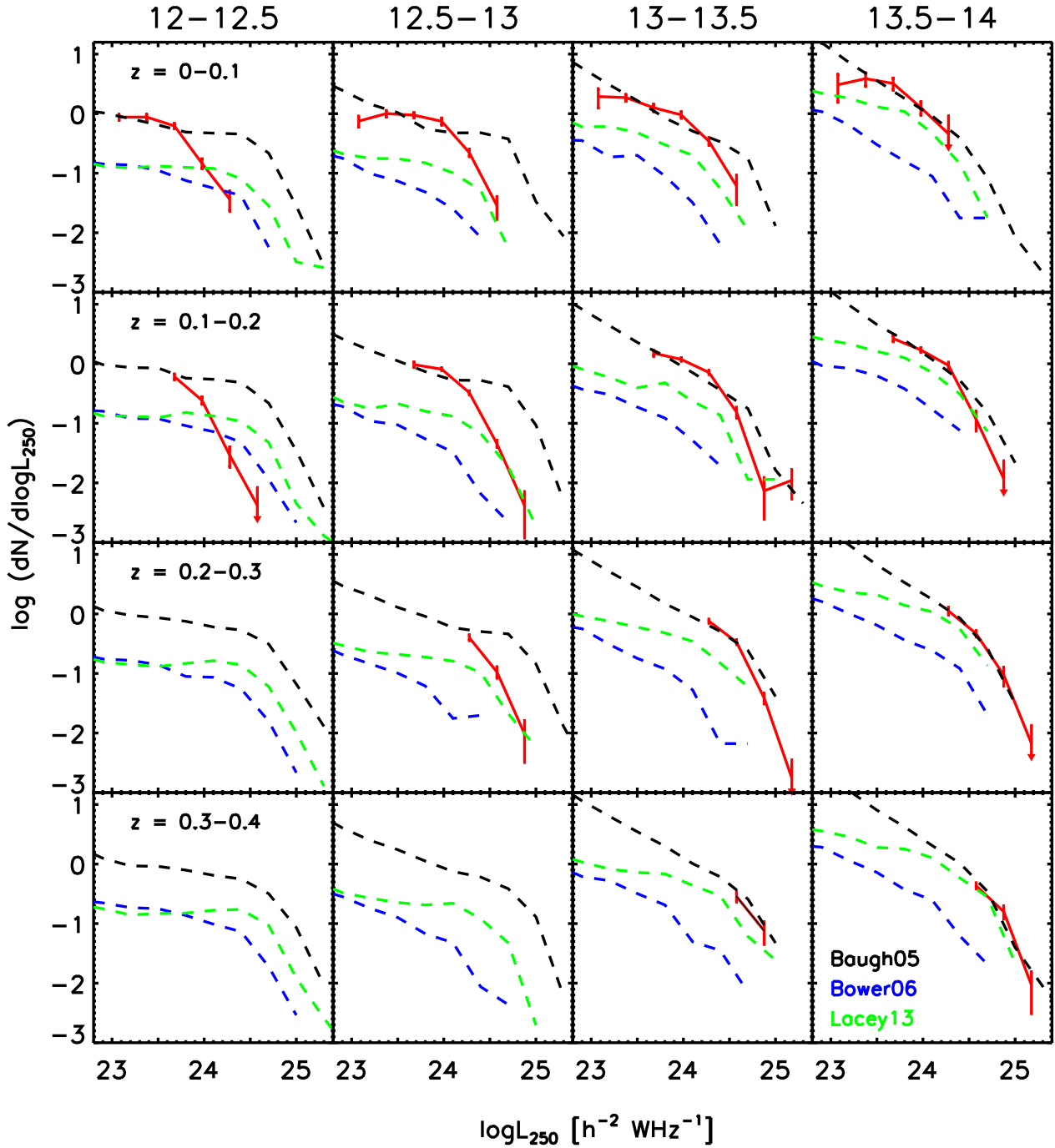}
\caption{Comparison between the observational group CLFs and galaxy
  formation model predictions. Red solid curves show our measurements
  for \HATLAS groupswith optical multiplicity $n\geq 2$. Dashed curves
  show predictions for different \GALFORM models, as indicated in the
  key.}
\label{fig:comSAM}
\ec
\end{figure*}

Semi-analytical modelling of galaxy formation in the $\Lambda$CDM
framework has been proven very powerful in reproducing many observed
properties of galaxies and their evolution
\citep[e.g.][]{Cole2000,Baugh2005, Bower2006,Croton2006, Delucia2007,
Guo2011b}.  However, until recently there has been only limited
theoretical work combining galaxy formation models with modelling of
the far-IR emission in a cosmological context
\citep{Granato2000,Devriendt2000,Baugh2005,Lacey2008,Lacey2010,
Somerville2012}. Here we compare our measurements of the far-IR
luminosity function in groups to predictions from the \GALFORM
semi-analytical model \citep{Cole2000}. \GALFORM incorporates a
treatment both of the absorption of starlight by dust in galaxies and
of the far-IR emission by the dust heated in this way (see
\citealt{Lacey2011} and \citealt{Lacey2013} for more details). We show
predictions from three different versions of the \GALFORM model,
namely \citet{Baugh2005}, \citet{Bower2006} and
\citet{Lacey2013}. These models differ in several ways. The
\citet{Baugh2005} model has a top-heavy initial mass function (IMF)
for stars formed in starbursts, which was introduced in order to
reproduce the number counts and redshift distribution of the faint
sub-mm galaxy population detected at 850$\mum$. The \citet{Bower2006}
model has a single IMF and includes AGN feedback, but does not
reproduce the sub-mm galaxies. The \citet{Lacey2013} model includes
both AGN feedback and a top-heavy IMF in starbursts (though less top-heavy than that used in the Baugh et al. model). It matches the
number counts and redshift distribution at 850$\mum$, and was also
adjusted to approximately fit the observed number counts in the 250,
350 and 500$\mum$ bands. None of these models had their parameters
adjusted with reference to any observed properties of galaxy groups,
so these are ``blind'' predictions. Rather than identify galaxy groups
in the \GALFORM simulations in the same way as   stop
done for the \GAMA
group catalogue, we simply plot CLFs for virialized dark matter halos
of different masses.

The results are presented in Fig. ~\ref{fig:comSAM}. The red curves
with error bars are our observational results for groups with
multiplicity $n\geq$ 2. The dashed curves in different colours show
the predictions for the three different \GALFORM models. We first
emphasize that all the models predict that the amplitude of the CLF
(i.e. the number of galaxies per group) increases with group mass and
with redshift, in qualitative agreement with our observational
measurements.  In general, the predictions from the \cite{Baugh2005}
model are in best agreement with our measured far-IR CLFs, though this
model still predicts too many galaxies with high far-IR luminosities
in lower-mass groups.  The \cite{Bower2006} model underestimates the
abundance of far-IR galaxies over the whole range of group mass and
redshift studied here, generally by a large factor. The predictions of
the \citet{Lacey2013} model lie between those of the other two
models. None of the models reproduces the trend of
characteristic far-IR luminosity strongly increasing with halo mass
that we see in the observations. We conclude that observations of the
far-IR CLFs of groups can put stringent new constraints on galaxy
formation models, which are complementary to the standard
observational constraints (such as from galaxy luminosity functions)
that are typically used. In particular, the far-IR CLFs of groups
tightly constrain how star formation in galaxies depends on the host
halo mass, which in turn puts constraints on physical processes in
galaxy formation models such as gas cooling, stripping  and feedback from
supernovae and AGN. We plan to explore these constraints in more
detail in a future paper.

\section{Discussion and Conclusions}
\label{sec:conc}
Observations at far-IR wavelengths are an essential complement to the
traditional UV and optical tracers of star formation. We have combined
far-IR data from the \HATLAS survey with the galaxy group catalogue
from the \GAMA optical spectroscopic survey to study the far-IR
luminosity functions of galaxies in different group environments and
at different redshifts. We use a sample of 10.5k galaxies from the
\HATLAS survey, flux limited at 250$\mum$ with $S(250) >$ 32 \mJy, and
matched to $r$-band selected galaxies with $r<19.4$ in the \GAMA
spectroscopic survey, together with a catalogue of 10.7k \GAMA groups
in the same region.  We have used two independent methods to estimate
the conditional far-IR luminosity functions of groups. One is to
directly identify the group membership of each far-IR source by
matching to its optical counterpart. The other is to count the average
excess number of far-IR galaxies within the optically-estimated radius
of each group. The measured far-IR luminosity functions as a function
of group mass and redshift are consistent between these two methods,
but the results from the direct method are less noisy, so we use the
direct method for most of our analysis. We find that the far-IR
luminosity functions are insensitive to the group optical multiplicity
for a given group mass and redshift. We have measured average far-IR
luminosity functions in bins of mass and redshift over a range of
$10^{12} < M_{\rm h} < 10^{14} \hMsol$ in group mass and a range of $0<z<0.4$
in redshift, probing galaxy IR luminosities $L_{\rm IR} > 2\times 10^9
\hLsol$. 

We find that the far-IR luminosity functions of groups are well fitted by
a modified Schechter function, as previously found for the field
population. We find that the characteristic far-IR luminosity
$L_{\ast}$ of galaxies in groups increases with the group mass below
$10^{13}{h^{-1}}\rm M_{\odot}$, while at higher masses it flattens
or turns over. The redshift dependence of $L_{\ast}$ is a strong
function of group mass. For very massive systems, $L_{\ast}$ at $z\sim
0.3$ is 2.5 times larger than at $z \sim 0$, while this difference
between high and low redshifts nearly vanishes for group masses below
$10^{12.5}{h^{-1}}\rm M_{\odot}$.  By integrating over the far-IR
luminosity function of galaxies in groups, we calculate the ratio
$L_{\rm IR}/M_{\rm h}$ of total IR luminosity to group mass. We find that this
ratio is a decreasing function of group mass and an increasing
function of redshift, being fit by $L_{\rm IR}/M_{\rm h} \propto M_{\rm h}^{-0.65}
(1+z)^{5}$. We estimate that for $z<0.2$, around 70\% of the total
far-IR luminosity density is contributed by galaxies in halos more
massive than $10^{12} {\rm \hMsol}$.


We also use our \HATLAS/\GAMA galaxy sample to measure the relation
between far-IR/$r$-band colour and $r$-band absolute magnitude in the
field and in groups of different mass. For $z<0.05$, we find that for far-IR detected galaxies this
relation is independent of group mass over the whole range $10^{12} <
M_{\rm h} < 10^{14} \hMsol$, and the same as that in the field.  
Similarly, we find the average derived sSFR 
has only a weak dependence on galaxy stellar mass, and the sSFR vs. stellar mass relation 
is indistinguishable in different environments. Again, this result applies to galaxies in our sample with detectable far-IR emission. This result
is consistent with most previous studies of the dependence of star
formation rates of actively star-forming galaxies on environment using
UV and optical tracers \citep[e.g.][]{Balogh2004, Tanaka2004,
Weinmann2006, Peng2010, McGee2011}. For $z>0.05$ we find that no firm
conclusions can be drawn about the far-IR/optical colour-magnitude
relation from this sample, due to the far-IR flux limit.

We compared our results on the far-IR luminosity functions of groups
to three different semi-analytical galaxy formation models which have
already proven successful in producing many other galaxy properties
both at high and low redshifts. All these models qualitatively
reproduced the trend of the characteristic far-IR luminosity
$L_{\ast}$ increasing with group mass and redshift. However, none of
them were able to reproduce the observed conditional far-IR luminosity
functions in detail. This impli es some deficiency in the way physical
processes such as gas cooling, star formation and feedback are
calculated in current galaxy formation models, but also demonstrates
the potential for using such observations to distinguish between
different models. Our comparison with the models assumed that the
galaxy groups identified in the \GAMA survey correspond closely in
both galaxy membership and total mass to the dark matter halos in the
theoretical galaxy formation models. In future work, we plan to test
these assumptions by constructing mock galaxy catalogues from the
models and applying the same algorithms for identifying groups and
measuring their far-IR luminosity functions as for the observations.


The analysis in this paper is based entirely on far-IR luminosities,
which trace the dust-obscured component of galaxy SFRs, while the
unobscured component of galaxy SFRs is traced by their far-UV
luminosities. Most of the area covered by \HATLAS PhaseI and \GAMA
surveys also has far-UV imaging from \GALEX. In a future paper, we
plan to combine the far-IR and far-UV data from \HATLAS and \GAMA to
estimate total galaxy SFRs free from biases due to dust obscuration,
and use these to study their dependence on environment and redshift,
in a similar way as done here for the far-IR luminosities.

\section*{Acknowledgments}

The authors thank Michal Michalowski and Jochen Liske for useful
comments.  QG acknowledges support from a Newton International
Fellowship, the NSFC grants (Nos 11143005 and No.11133003 ) and
the Strategic Priority Research Program “The Emergence of Cosmological
Structure” of the Chinese Academy of Sciences (No. XDB09000000). PN acknowledges the support of the Royal Society through
the award of a University Research Fellowship and the European
Research Council, through receipt of a Starting Grant
(DEGAS-259586). CSF acknowledges a Royal Society Wolfson Research
Grant Award.  This work was supported in part by the Science and
Technology Facilities Council rolling grant ST/F001166/1 to the ICC.
Calculations were partly performed on the ICC Cosmology Machine, which
is part of the DiRAC Facility jointly funded by STFC and Durham
University.

The Herschel-ATLAS is a project with Herschel, which is an ESA space
observatory with science instruments provided by European-led
Principal Investigator consortia and with important participation from
NASA. The H-ATLAS website is http://www.h-atlas.org/

GAMA is a joint European-Australian project based around a
spectroscopic campaign using the Anglo- Australian Telescope. The GAMA
input catalogue is based on data taken from the Sloan Digital Sky
Survey and the UKIRT Infrared Deep Sky Survey. Complementary imaging
of the GAMA regions is being obtained by a number of independent
survey programs including GALEX MIS, VST KiDS, VISTA VIKING, WISE,
Herschel-ATLAS, GMRT and ASKAP providing UV to radio coverage. GAMA is
funded by the STFC (UK), the ARC (Australia), the AAO, and the
participating institutions. The GAMA website is:
http://www.gama-survey.org/

The HATLAS-ATLAS and GAMA data will become public in the future. Details
can be found on their websites. For more information about the model galaxy catalogues please  contact the 
corresponding authors.

\bibliographystyle{mn2e}

\setlength{\bibhang}{2.0em}
\setlength\labelwidth{0.0em}

\bibliography{draft}

\appendix

\section{Group CLFs with the stacking method}
\label{sec:stacking-method}

In this Appendix we give a few more details about the stacking method
and the results obtained using it.

\begin{figure*}
\bc
\hspace{-0.6cm}
\includegraphics[width=18cm, bb= 28 40 600 622]{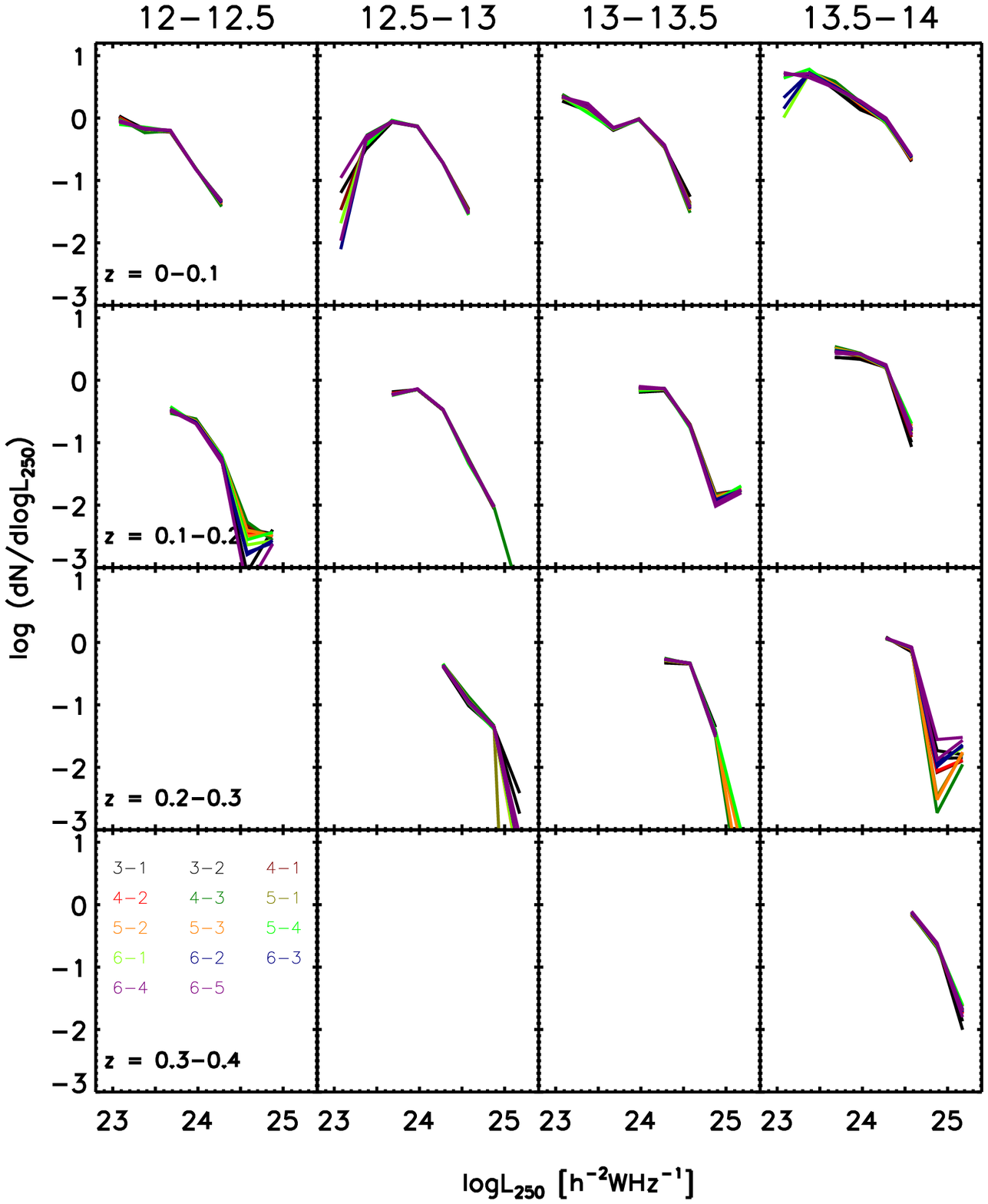}
\caption{Conditional far-IR luminosity functions of galaxy groups
  measured with the stacking method for different choices of local
  background subtraction. The background number density around each
  group is estimated in an annulus. The different choices for the outer
  inner radii of these annuli in units of the group radius
  $R_{100}$ are indicated in the bottom left panel (see text for
  details.)  }
\label{fig:background}
\ec
\end{figure*}

Our stacking method for measuring the CLF in groups involves
subtracting the estimated local background density of galaxies from
the total projected counts, which means that our results might be affected by
how we measure this background. We estimate the local background
density within an annulus around each group. If we make the radius of
this annulus too small, then our background estimate may include
galaxies associated with the group due to clustering, causing us to
overestimate the background and so underestimate the CLF of group
members. On the other hand, if we make the annulus too large, we may
fail to properly subtract the effect of foreground or background
structures, again causing an error in the measured CLF.  Here we test
the effect on the measured CLFs of varying the inner and outer radii
of the annuli, where these are taken to be fixed multiples
of the group radius $R_{100}$. The results are shown in
Fig. \ref{fig:background}, split by group mass and redshift, with radius 
range of the annulus in units of R$_{100}$ for each coloured line shown by the key in the bottom
left panel. This figure shows that our measured CLFs are not very
sensitive to the choice of annulus for the background subtraction.

\begin{figure*}
\bc
\hspace{-0.6cm}
\includegraphics[width=18cm, bb= 28 40 600 622]{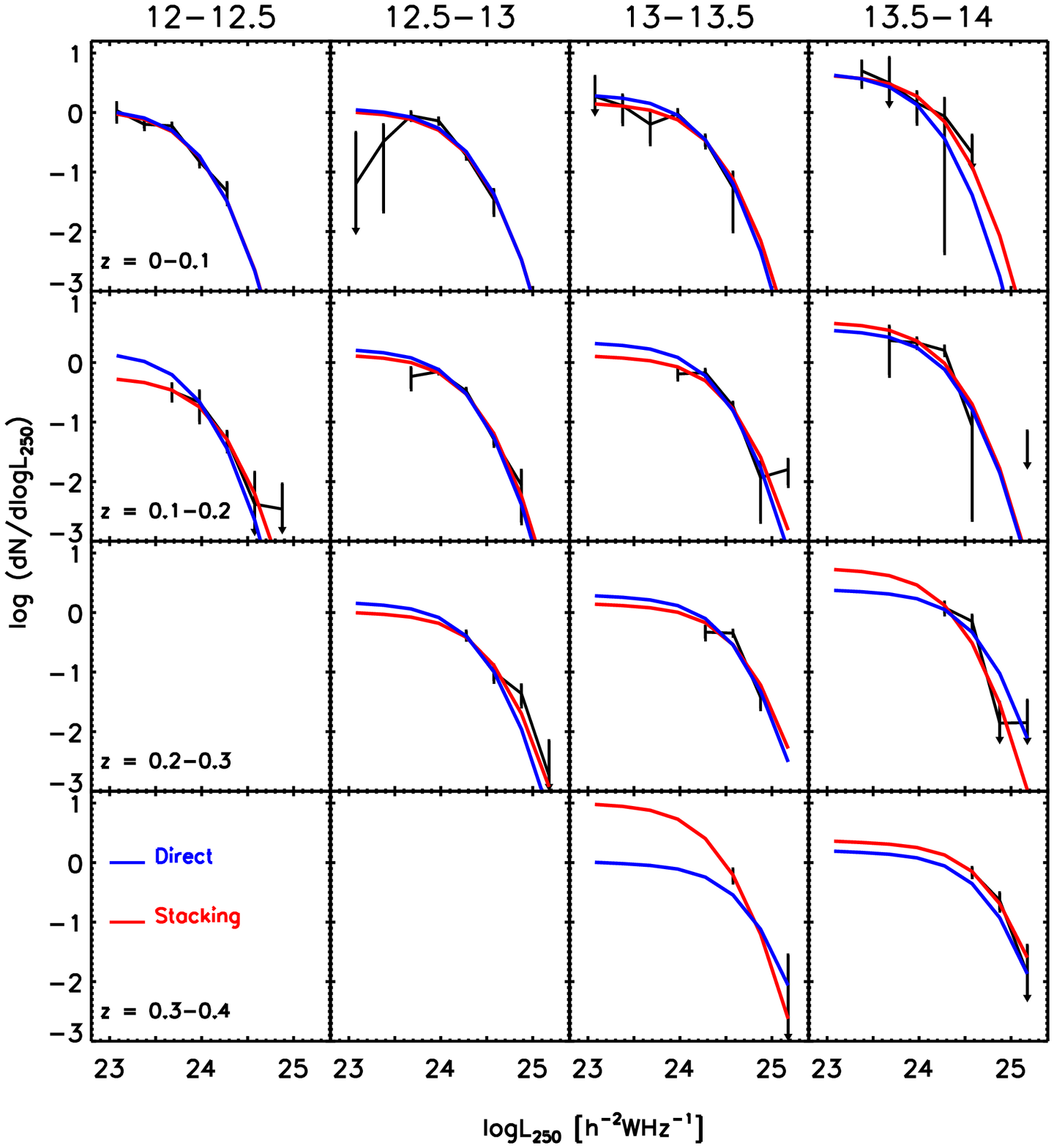}
\caption{Conditional far-IR luminosity functions of galaxies in groups
  measured using the stacking method. The black lines with errorbars
  are the direct measurement, and the red curves are the analytic fits
  to these. The blue curves replicate the analytic fits from the
  direct method. Note that in most panels, the measurements are quite
  noisy, and these blue curves fits the stacking method measurements
  reasonably well. The parameters $\alpha$ and $\sigma$ in
  eqn (\ref{eq:saunders}) are fixed at 1.06 and 0.30, as measured from
  the field luminosity function. 
The redshift ranges are indicated
  in the first column and the group mass ranges are indicated on the
  top of each panel in the first row.  }
\label{fig:stackfitting}
\ec
\end{figure*}

In Fig. \ref{fig:stackfitting} we show the group CLFs measured using
the stacking method (in black). These CLFs are noisier than those
measured using the direct method (shown in
Fig. \ref{fig:clf_fitting}). We also show the analytic fits using
eqn (\ref{eq:saunders}) to the measured CLFs from the stacking method
(in red) and from the direct method (in blue). It can be seen that the
fits from the stacking method agree reasonably well with those from
the direct method in all group mass and redshift ranges.

\end{document}